\documentstyle[aps,prd,preprint,epsfig,tighten]{revtex}
\begin{document}
\draft
\title{		Four-neutrino oscillation solutions \\ 
		of the atmospheric neutrino anomaly}
\author{ 	G.L.\ Fogli, E.\ Lisi, and A.\ Marrone}
\address{	Dipartimento di Fisica and Sezione INFN di Bari\\
             	Via Amendola 173, I-70126 Bari, Italy \\ }
\maketitle
\begin{abstract}
In the context of neutrino scenarios characterized by four (three active plus
one sterile) neutrino species and by mass spectra with two separated doublets,
we analyze solutions to the atmospheric neutrino anomaly which smoothly
interpolate between $\nu_\mu\to\nu_\tau$ and $\nu_\mu\to\nu_s$ oscillations. We
show that, although the Super-Kamiokande data disfavor the {\em pure\/}
$\nu_\mu\to\nu_s$ channel, they do not exclude its occurrence,  with sizable
amplitude, {\em in addition\/} to the $\nu_\mu\to\nu_\tau$   channel. High
energy muon data appear to be crucial in assessing the relative amplitude of
active and sterile neutrino oscillations. It is also qualitatively shown that
such atmospheric $\nu$ solutions are compatible with analogous solutions to the
solar neutrino problem, which involve oscillations of $\nu_e$ in both sterile
and active states.
\end{abstract}
\medskip

\section{Introduction}

Four-neutrino ($4\nu$) models, involving the three known active states
$(\nu_e,\nu_\mu,\nu_\tau)$ plus one hypothetical sterile state ($\nu_s$), are
being intensively studied in the context of the current neutrino 
phenomenology, since they can accommodate (through three independent mass
square differences) the three sources of evidence for $\nu$ flavor oscillations
coming from atmospheric, solar, and accelerator neutrino experiments. In
particular, $4\nu$ spectra with mass eigenstates  organized in two doublets
(2+2 models) seem to be favored by world  neutrino data \cite{Four}. Updated
reviews on such $4\nu$ models and on their phenomenology  can be found, e.g.,
in \cite{Bi99,Do00,Ba00}, to which the reader is referred   for an extended
bibliography and for details that are not repeated in this work.

According to the conventional wisdom,  2+2 models are often assumed to imply,
for the solar and atmospheric oscillation channels, either 
\begin{mathletters}
\begin{eqnarray}
\nu_\mu &\to& \nu_\tau\;\; ({\rm atmospheric})\ ,\label{mutau}\\
\nu_e &\to& \nu_s\;\; ({\rm solar})\ ,\label{es}
\end{eqnarray}
\end{mathletters}
or
\begin{mathletters}
\begin{eqnarray}
\nu_\mu &\to& \nu_s\;\; ({\rm atmospheric})\ ,\label{mus}\\
\nu_e &\to& \nu_\tau\;\; ({\rm solar})\ , \label{etau}
\end{eqnarray}
\end{mathletters}
in addition to inter-doublet $\nu_\mu\to\nu_e$ oscillations with small
amplitude, required to explain the Liquid Scintillator Neutrino Detector (LSND)
signal \cite{LSND,LS2K}.

Such simplifying assumptions are challenged by the recent Super-Kamiokande (SK)
observations,   which tend to disfavor (dominant) oscillations into $\nu_s$ of
both  atmospheric muon neutrinos (at $\sim 99\%$ C.L. \cite{So2K,SK2K}) and 
solar electron neutrinos (at $\sim 95\%$ C.L.,  in combination with world solar
$\nu$ data \cite{Su2K}). The MACRO experiment also disfavors  atmospheric
$\nu_\mu\to\nu_s$ oscillations (at $\sim 95\%$ C.L.\ \cite{Ba2K}). However, it
seems premature to rule out the sterile neutrino hypothesis on the basis of the
present information only \cite{Go2K,Fate}:  on the one hand, past experience on
global data fits has shown that, sometimes,  hypotheses rejected (accepted) at
95\% or 99\% C.L.\  can  surprisingly revive (die); on the other hand, the
underlying $4\nu$ oscillation pattern might be more complicated
\cite{Do00,Pena,Pe2K} than assumed in Eqs.~(\ref{mutau},\ref{es})  or
(\ref{mus},\ref{etau}).  In particular, instead of having decoupled active and
sterile oscillation channels, one might have mixed (active+sterile) flavor
transitions of the kind \cite{Li2K}
\begin{mathletters}
\begin{eqnarray}
\nu_\mu &\to& \nu_+\;\; ({\rm atmospheric})\ ,\label{mu+}\\
\nu_e &\to& \nu_-\;\; ({\rm solar})\ ,\label{e-}
\end{eqnarray}
\end{mathletters}
where  the states $\nu_\pm$, as discussed later in more detail, represent
linear (orthogonal)  combinations of $\nu_\tau$ and $\nu_s$ through a mixing
angle $\xi$,
\begin{mathletters}
\begin{eqnarray}
\nu_+ &=& +\cos\xi\,\nu_\tau + \sin\xi\,\nu_s\ ,\label{nu+}\\
\nu_-&=& -\sin\xi\, \nu_\tau + \cos\xi\, \nu_s\ .
\label{nu-}
\end{eqnarray}
\end{mathletters}
The oscillation modes~(\ref{mu+},\ref{e-}) represent generalizations of both
modes (\ref{mutau},\ref{es}) and (\ref{mus},\ref{etau}), to which they reduce
for $\sin\xi=0$ and 1, respectively. For generic values of $\sin\xi$, the
final  states in atmospheric $\nu_\mu$ and solar $\nu_e$ flavor transitions are
linear combinations of $\nu_s$ and $\nu_\tau$, with coefficients to be
constrained by experiments.

In this work, we quantitatively study atmospheric four-neutrino oscillations in
the context of 2+2 spectra,  for unconstrained values of
$\sin\xi=\langle\nu_+|\nu_s\rangle$. It is shown that the state $\nu_+$ (into
which $\nu_\mu$ oscillates) can have a sizable $\nu_s$ component, which is
testable through high-energy atmospheric muon data.  Our atmospheric $\nu$
results are also qualitatively compared with recent $4\nu$ solutions to the
solar neutrino problem \cite{Pena,Pe2K}, which are compatible with a large
$\nu_s$ component of $\nu_-$. It is shown that both solar and atmospheric
neutrino oscillations are consistent with sizable flavor transitions to a
sterile state.

The structure of this paper is as follows. In Sec.~II we introduce the
theoretical $4\nu$ framework, and in Sec.~III we show the corresponding
graphical representation of the parameter space. Section~IV reports a
preliminary discussion of the SK experimental information, while Sec.~V is
devoted to a thorough SK atmospheric $\nu$ data analysis, from which we derive
constraints on $4\nu$ mass-mixing parameters. Their consistency with solar
neutrino data is discussed in Sec.~VI.  Remarks on alternative $4\nu$ scenarios
are made in Sec.~VII.  Finally, conclusions and perspectives are reported in
Sec.~VIII.

While this  work was being completed, our attention was brought to the paper
\cite{Ya00}, where $4\nu$ solutions to the atmospheric neutrino anomaly have
also been worked out. The $4\nu$ scenario considered in \cite{Ya00}  contains
two  additional parameters as compared with ours (a mixing angle $s_{23}$ and a
CP violation phase $\delta_1$ \cite{Ya00}), and most of the results shown 
refer to such parameters, so that a direct comparison with our results is
difficult. However, we agree with \cite{Ya00} on some general features, namely,
that atmospheric $\nu_\mu$ can have a large transition amplitude to $\nu_s$ (in
addition to $\nu_\mu\to\nu_\tau$), and that such transitions are consistent
with solar neutrino data.

\section{Theoretical framework}

In this section we define the theoretical $4\nu$ framework and the notation 
used in our analysis. We also motivate a few data-driven assumptions, which
considerably simplify both the analysis and the understanding of  the solar and
atmospheric $\nu$ oscillations.

\subsection{General conventions}

We order the neutrino flavor and mass eigenstates in column vectors as
\begin{eqnarray}
\nu_\alpha &=& (\nu_e,\nu_s,\nu_\mu,\nu_\tau)^T\ ,
\label{nualpha}\\
\nu_i&=&(\nu_1,\nu_2,\nu_3,\nu_4)^T\ ,
\label{nui}
\end{eqnarray}
respectively ($T=$~transpose).  Such vectors are related by a unitary mixing
matrix $U$,
\begin{equation}
\nu_\alpha = U_{\alpha i}\nu_i\ .
\label{Ualphai}
\end{equation}

The matrix of squared neutrino masses (in the $\nu_i$ basis) is defined as
\begin{equation}
{\cal M} = {\rm diag}\,(m^2_1,m^2_2,m^2_3,m^2_4) + \lambda^2 \mathbf{1}\ ,
\label{calM}
\end{equation}
where the  term $\lambda^2\mathbf{1}$ reminds that the overall mass scale
$\lambda$ is unconstrained,  as far as oscillation experiments  are concerned.
We define three independent mass square differences, phenomenologically related
to the evidence for oscillations coming from solar, atmospheric, and LSND
experiments,
\begin{mathletters}
\begin{eqnarray}
\delta m^2 &=& m^2_2-m^2_1\;\, ({\rm solar})\ ,\label{deltam2}\\
m^2 &=& m^2_4-m^2_3\;\; ({\rm atmos.})\ ,\label{m2}\\
M^2 &=& m^2_{3}-m^2_{2}\;\;({\rm LSND})\label {M^2}\ .
\end{eqnarray}
\end{mathletters}
With the above definitions, the neutrino evolution equation in the flavor
basis,
\begin{equation}
i\frac{d}{dx}\,\nu_\alpha = {\cal H}\,\nu_\alpha\ ,
\label{iH}
\end{equation}
is governed by the Hamiltonian
\begin{equation}
{\cal H}=\frac{1}{2E}\,U{\cal M}\,U^\dagger + 
\sqrt{2}\,G_F\,{\rm diag}\,\left(N_e,{\textstyle\frac{1}{2}}N_n,0,0\right)\ ,
\label{H}
\end{equation}
where the first (kinematical) term \cite{Pont}  depends on the neutrino energy
$E$, and the second (dynamical) term \cite{MSWs} depends on the neutrino weak
interactions with background electrons (with density $N_e$) and neutrons (with
density $N_n$) in matter.

\subsection{Approximations concerning $4\nu$ masses and mixing}

Motivated by the current $4\nu$  oscillation phenomenology (see
\cite{Bi99,Do00,Ba00} and references therein),  we assume that the mass
eigenstates are organized in  two doublets  (2+2 spectrum),
\begin{equation}
m_1\simeq m_2 < m_3\simeq m_4\ .
\end{equation}
with three widely different mass squared differences,
\begin{equation}
\delta m^2 \ll m^2 \ll M^2\ .
\label{gerarchia}
\end{equation}

Figure~1 shows our reference mass spectrum, with a lower ``solar neutrino
doublet'' $(\nu_1,\nu_2)$, and an upper ``atmospheric neutrino doublet''
$(\nu_3,\nu_4)$, separated by a relatively large LSND mass gap. Other
phenomenologically allowed 2+2 spectra can be obtained by interchanging the two
doublets ($M^2\to -M^2$), as well as the two states in a doublet ($\delta
m^2\to -\delta m^2$, or $m^2\to -m^2$, or both). We anticipate that, under the
approximations discussed below,  the oscillation physics in such alternative
2+2 spectra is equivalent (under appropriate changes of variables) to that in
the spectrum of Fig.~1. Therefore, our reference choice in Fig.~1 does not
represent a limitation, as far as 2+2 spectra are considered.%
\footnote{Although our work is focussed on 2+2 spectra, we will also briefly
comments upon alternative 3+1 spectra (a triplet plus a loner state)   in
Sec.~VII. In fact, the standard arguments against 3+1 models (see, e.g.,
\protect\cite{Four}) appear now to be somewhat  less compelling
\protect\cite{Fate,Sm2K} in light of the most recent LSND data
\protect\cite{LS2K}. }
We also anticipate that, under the same approximations, CP violation effects
turn out to be unobservably small in current experiments, so that the mixing
matrix $U$ can be effectively taken as real,
\begin{equation}
U^\dagger\simeq U^T
\label{Ureal}\ .
\end{equation}

Further assumptions about the mixing matrix $U$ can be adopted  by comparing 
positive disappearance results at the mass scale $\delta m^2$ (solar
$\nu_e\to\nu_e$) and $m^2$ (atmospheric $\nu_\mu\to\nu_\mu$) with negative
disappearance results at the mass scale $M^2$ in short-baseline accelerators 
(no $\nu_\mu\to\nu_\mu$) and reactors (no $\nu_e\to\nu_e$) (see, e.g.,
\cite{Do00}). Such results imply that the $\nu_\mu$ component of the
atmospheric doublet $(\nu_3,\nu_4)$  must be large to explain the atmospheric
$\nu_\mu$ anomaly,  but must be small in the other doublet $(\nu_1,\nu_2)$, in
order to avoid large $M^2$-driven $\nu_\mu\to\nu_\mu$ oscillations at
short-baseline accelerators. Analogously, the $\nu_e$ component of the solar
doublet $(\nu_1,\nu_2)$ must be large to explain the solar $\nu_e$ deficit, but
it must be small in the other doublet $(\nu_3,\nu_4)$, in order to avoid large
$M^2$-driven  $\nu_e\to\nu_e$ oscillations at reactors. Therefore, one can take
\begin{eqnarray}
U^2_{\mu1}+U^2_{\mu2} &\simeq& 0
\label{Umu12}\ ,\\ 
U^2_{e3}+U^2_{e4} &\simeq& 0\ .
\label{Ue34}
\end{eqnarray}

Some remarks on the above two assumptions are in order. In $4\nu$ models
embedding LSND, Eqs.~(\ref{Umu12}) and (\ref{Ue34}) must  be slightly
violated---at the few permill level---in order to get a nonzero value for the
small (few permill) LSND $\nu_\mu\to\nu_e$ oscillation amplitude, given by
$A_{\rm LSND}=4|U_{\mu3}U_{e3}+U_{\mu4}U_{e4}|^2
=4|U_{\mu1}U_{e1}+U_{\mu2}U_{e2}|^2$ \cite{Bi99,Do00,Ba00}.  A few permill
violation makes little difference in the analysis of current solar and
atmospheric neutrino data, and will be neglected in the following. One cannot
exclude, however, more general scenarios  in which $U^2_{\mu 1}+U^2_{\mu 2}$ is
taken close to the weakest upper limits allowed by accelerators (of the order
of ten percent \cite{CDHS})  and can thus play a nonnegligible phenomenological
role \cite{Do00,Ya00}. Such scenarios are beyond the scope of the present work.

The simplifying assumptions (\ref{Umu12}) and (\ref{Ue34}) lead to a specific
texture for the matrix $U$,
\begin{equation}
U\simeq 
\left(
\begin{array}{cccc}
U_{e1} & U_{e2} &  0 & 0 \\
U_{s1} &U_{s2} &U_{s3} &U_{s4} \\
0 & 0 & U_{\mu3} & U_{\mu4} \\
U_{\tau1} &U_{\tau2} &U_{\tau3} &U_{\tau4}
\end{array}
\right)\ ,
\label{Usimeq}
\end{equation}
with the following implications. The state $\nu_\mu$ ($\nu_e$) must be a linear
combination of $\nu_3$ and $\nu_4$ ($\nu_1$ and $\nu_2$),
\begin{eqnarray}
\nu_\mu &=& +c_\psi \nu_3 + s_\psi\nu_4\ ,\label{numu}\\
\nu_e&=& +c_\omega \nu_1 + s_\omega \nu_2\ ,
\label{nue}
\end{eqnarray}
where $c=\cos$, $s=\sin$, and we have introduced two mixing angles $\psi$ and
$\omega$ (ranging in $[0,\pi/2]$). The corresponding orthogonal combinations
$\nu_+$  and $\nu_-$, defined as
\begin{eqnarray}
\nu_+ &=& -s_\psi \nu_3 + c_\psi\nu_4\ ,\label{nu++}\\
\nu_-&=& -s_\omega \nu_1 + c_\omega \nu_2\ ,
\label{nu--}
\end{eqnarray}
must  be a linear combination of the remaining flavor states $\nu_\tau$ and
$\nu_s$,
\begin{eqnarray}
\nu_+ &=& +c_\xi\nu_\tau + s_\xi\nu_s\ ,\label{nutaus1}\\
\nu_-&=& -s_\xi \nu_\tau + c_\xi \nu_s\ ,
\label{nutaus2}
\end{eqnarray}
where $\xi$ is the third (and last) mixing angle needed to parametrize a
unitary matrix $U$ of the form (\ref{Usimeq}). In terms of $(\omega,\psi,\xi)$,
the matrix $U$ reads explicitly
\begin{equation}
U=
\left(
\begin{array}{cccc}
c_\omega & s_\omega & 0 & 0 \\
-s_\omega c_\xi & c_\omega c_\xi & -s_\psi s_\xi & c_\psi s_\xi \\
0 & 0 & c_\psi & s_\psi  \\
s_\omega s_\xi & -c_\omega s_\xi & -s_\psi c_\xi & c_\psi c_\xi
\end{array}
\right)\ .
\label{Uangles}
\end{equation}

By defining the rotation matrices 
\begin{equation}
U_\xi = 
\left(
\begin{array}{cccc}
1 & 0 & 0 & 0 \\
0 & c_\xi & 0 & s_\xi \\
0 & 0 & 1 & 0 \\
0 & -s_\xi & 0 & c_\xi 
\end{array}
\right)\ ,
\label{Uxi}
\end{equation}
\begin{equation}
U_\psi = 
\left(
\begin{array}{cccc}
1 & 0 & 0 & 0 \\
0 & 1 & 0 & 0 \\
0 & 0 & c_\psi &  s_\psi \\
0 & 0 & -s_\psi & c_\psi
\end{array}
\right)\ ,
\label{Upsi}
\end{equation}
and
\begin{equation}
U_\omega = 
\left(
\begin{array}{cccc}
c_\omega & s_\omega & 0 & 0\\
-s_\omega & c_\omega & 0 & 0 \\
0 & 0 & 1 & 0 \\
0 & 0 & 0 & 1
\end{array}
\right)\ ,
\label{Uomega}
\end{equation}
the mixing matrix $U$ can also be written as
\begin{equation}
U=U_\xi\, U_{\psi\omega}\ ,
\label{Uproduct}
\end{equation}
where the commutation property $[U_\omega,U_\psi]=0$ has been used:
\begin{equation}
U_{\psi\omega} \equiv U_\psi\, U_\omega = U_\omega\, U_\psi\ .
\label{Upsiomega} 
\end{equation}

Finally, it is useful to  introduce a new flavor basis $\nu'_\alpha$ defined as
\begin{equation}
\nu'_\alpha = (\nu_e,\nu_-,\nu_\mu,\nu_+)^T\ ,
\label{nu'alpha}
\end{equation}
related to the standard flavor basis by a $\xi$-rotation,
\begin{equation}
\nu'_\alpha = U_\xi^T \nu_\alpha\ .
\end{equation}
In such  basis, the neutrino evolution Hamiltonian reads
\begin{equation}
{\cal H'}=\frac{1}{2E}\,U_{\psi\omega}^{\phantom{T}}
{\cal M}\,U_{\psi\omega}^T + \sqrt{2}\,G_F\,U^T_\xi\,
{\rm diag}\left(N_e,{\textstyle\frac{1}{2}}N_n,0,0\right)\,U_\xi\ .
\label{H'}
\end{equation}
In the next subsection, it will be shown how the above $4\nu$ Hamiltonian can
be effectively decoupled in two (solar and atmospheric) $2\nu$
sub-Hamiltonians.

Effective  Hamiltonians for mixed active-sterile oscillations in matter have
been recently discussed also in Refs.~\cite{Do00,Ya00}. We can make contact
with the $\phi_{ij}$ notation of \cite{Do00} through the  identifications
$\omega=\phi_{12}$, $\psi=\phi_{34}$, and $\xi=\phi_{24}$, valid under the
approximations $\phi_{23}\simeq 0$ [equivalent to Eq.~(\protect\ref{Umu12})]
and $\phi_{13}\simeq\phi_{14}\simeq 0$ [equivalent to
Eq.~(\protect\ref{Ue34})]. Analogously, we can make contact with the
$\theta_{ij}$ notation of \cite{Ya00} through the identifications
$\xi=\theta_{34}$ and $\psi=\theta_{24}$, valid under the approximation
$\theta_{23}\simeq 0$  [equivalent to Eq.~(\protect\ref{Umu12})].

\subsection{Approximations related to atmospheric $\nu$'s}

As far as atmospheric neutrinos are concerned, we neglect in first
approximation the small $(\nu_1,\nu_2)$ mass square difference $\delta m^2$,
and take the inter-doublet mass square difference $M^2$ very large. The matrix
${\cal M}$ in Eq.~(\ref{calM}) for atmospheric  neutrinos can then be written
as
\begin{equation}
{\cal M}_A \simeq {\rm diag}\,(-M^2,-M^2,0,m^2) \ .
\label{calMA}
\end{equation}

By inserting such ${\cal M}_A$ in Eq.~(\ref{H'}), one obtains that: $i)$ The
angle $\omega$ is rotated away; and $ii)$ as $M^2\to\infty$, the states
$(\nu_e,\nu_-)$ [or, equivalently, $(\nu_1,\nu_2)$] oscillate with very high
frequency ($\propto M^2$), and effectively decouple from the states
$(\nu_\mu,\nu_+)$. Therefore, only the evolution of  $(\nu_\mu,\nu_+)$  is
relevant,  
\begin{equation}
i\,\frac{d}{dx}\left(\begin{array}{c}\nu_\mu\\
\nu_+\end{array}\right) \simeq
\cal{H}'_A \left(\begin{array}{c}\nu_\mu\\
\nu_+\end{array}\right)\ ,
\end{equation}
the  Hamiltonian being given by
\begin{equation}
{\cal{H}}'_A = \frac{m^2}{4E}\left(\begin{array}{cc}c_{2\psi} & s_{2\psi}\\
s_{2\psi}& -c_{2\psi}\end{array}\right)+\sqrt{2}\,G_F\,
\left(\begin{array}{cc}0 & 0\\
0 & {\textstyle\frac{1}{2}}\, s^2_\xi\, N_n\end{array}\right)\ ,
\label{calHA}
\end{equation}
where we have subtracted an inessential term $m^2/4E$ from the diagonal elements
of ${\cal{H}}'_A $.

Equation~(\ref{calHA}) for atmospheric neutrinos is equivalent to the familiar
one describing the $\nu_\mu\to\nu_s$ Hamiltonian, provided that the usual mass
and mixing parameters are identified with $m^2$ and $\psi$, and that the
neutron density $N_n$ is multiplied by a factor $s^2_\xi$. For increasing
values of $s^2_\xi$, one gets a smooth interpolation from pure
$\nu_\mu\to\nu_\tau$ oscillations ($s^2_\xi=0$) to pure $\nu_\mu\to\nu_s$
oscillations ($s^2_\xi=1$), passing through mixed active-sterile transitions
$(0<s^2_\xi<1)$.

Equation~(\ref{calHA}), which has been derived for the reference $4\nu$
spectrum of Fig.~1,  trivially holds also for  for spectra with negative
$\delta m^2$ or negative $M^2$ (being independent of such mass parameters).
Concerning the case of negative $m^2$, one can easily prove (by swapping the
states $\nu_3$ and $\nu_4$), that it is equivalent to the case of positive
$m^2$, modulo the replacement $s_\psi \to c_\psi$ (and $\nu_+\to-\nu_+$),
which  corresponds to swap the first two octants of $\psi$, and thus does not
need a separate treatment (as far as $\psi\in [0,\pi/2]$). It follows that the
sign of $m^2$ is irrelevant at the octant boundary $(\psi=\pi/4$), namely, at
maximal atmospheric $\nu$ mixing.  For atmospheric antineutrinos, the same 
considerations as for neutrinos apply,  provided that the neutron density is
taken with opposite sign.

\subsection{Approximations related to solar $\nu$'s}

As far as solar neutrinos are concerned, the atmospheric neutrino doublet
$(\nu_3,\nu_4)$ can be seen as a couple of ``far'' mass eigenstates, separated
from $(\nu_1,\nu_2)$ by a large $M^2$ gap. At zeroth order in $\delta m^2/m^2$
and in $m^2/M^2$, the matrix ${\cal M}$ in Eq.~(\ref{calM}) for solar 
neutrinos can then be written as 
\begin{equation}
{\cal M}_S \simeq {\rm diag}\,(0,\delta m^2,M^2,M^2)\ .
\label{calMS}
\end{equation}

By inserting such ${\cal M}_S$ in Eq.~(\ref{H'}), one obtains that: $i)$ The
angle $\psi$ is rotated away; and $ii)$  as $M^2\to\infty$,  the states
$(\nu_\mu,\nu_+)$ [or, equivalently, $(\nu_3,\nu_4)$] oscillate with very high
frequency ($\propto M^2$), and effectively decouple from the states
$(\nu_e,\nu_-)$. Therefore, only the evolution of $(\nu_e,\nu_-)$  is relevant,
\begin{equation}
i\,\frac{d}{dx}\left(\begin{array}{c}\nu_e\\
\nu_-\end{array}\right)\simeq 
\cal{H}'_S \left(\begin{array}{c}\nu_e\\
\nu_-\end{array}\right)\ ,
\end{equation}
the Hamiltonian being given by
\begin{equation}
{\cal{H}}'_S = \frac{\delta m^2}{4E}
\left(\begin{array}{cc}c_{2\omega} & s_{2\omega}\\
s_{2\omega}& -c_{2\omega}\end{array}\right)+\sqrt{2}\,G_F\,
\left(\begin{array}{cc}N_e-{\textstyle\frac{1}{2}}\, c^2_\xi\, N_n & 0\\
0 & 0 \end{array}\right)\ ,
\label{calHS}
\end{equation}
where we have subtracted an inessential term  $(\delta
m^2/4E+\sqrt{2}\,G_F\,\frac{1}{2}\,c^2_\xi\, N_n)$ from the diagonal elements
of ${\cal{H}}'_S$.

Equation~(\ref{calHS}) for solar neutrinos is equivalent to the familiar one
describing two-family oscillations, provided that the usual mass and mixing
parameters are  identified with $\delta m^2$ and $\omega$, and that the
background fermion density is taken as $N_f=N_e-\frac{1}{2}c^2_\xi N_n$.  For
increasing values of $s^2_\xi$, one gets a smooth interpolation between pure
$\nu_e\to\nu_s$ oscillations ($s^2_\xi=0$ and $N_f=N_e-\frac{1}{2}N_n$)  and
pure $\nu_e\to\nu_\tau$ oscillations ($s^2_\xi=1$ and $N_f=N_e$), passing
through mixed  active-sterile transitions $(0<s^2_\xi<1)$.

Equation~(\ref{calHS}), which has been derived for the $4\nu$ reference
spectrum of Fig.~1,  trivially holds also for spectra with negative $m^2$ or
negative $M^2$ (being independent of such mass parameters). Concerning the case
of negative $\delta m^2$, one can easily prove  (by swapping the states $\nu_1$
and $\nu_2$), that it is equivalent to the case of positive $\delta m^2$,
modulo the replacement  $s_\omega \to c_\omega$ (and $\nu_-\to-\nu_-$), which
correspond  to swap the first two octants of $\omega$, and thus does not need a
separate treatment (as far as $\omega\in [0,\pi/2]$). It follows that the sign
of $\delta m^2$ is irrelevant at the octant boundary $(\omega=\pi/4$), namely,
at maximal solar $\nu$ mixing.

\subsection{Summary and remarks about the $4\nu$ framework}

We have introduced a 2+2 four-neutrino scenario, based on a few
phenomenological approximations, which embeds at the same time both active and
sterile oscillations of atmospheric and solar neutrinos.  The final states
$\nu_+$ and $\nu_-$  for atmospheric and solar flavor transitions
($\nu_\mu\to\nu_+$ and $\nu_e\to\nu_-$, respectively)  represent orthogonal
combinations of  $\nu_\tau$ and $\nu_s$ through a mixing angle $\xi$
[Eqs.~(\ref{nutaus1}) and  (\ref{nutaus2})].

The  angle $\xi$ governs the relative amount of $\nu_s$ and $\nu_\tau$ in the
states $\nu_\pm$, and thus also the amplitude of matter effects. When
$\xi\to0$, the state $\nu_+$ (oscillation partner of $\nu_\mu$)  is dominantly
a $\nu_\tau$, while the state $\nu_-$ (oscillation partner of $\nu_e$)  is
dominantly a $\nu_s$; and vice versa for $\xi\to \pi/2$. For $\xi=\pi/4$, both
solar and atmospheric neutrino oscillations are simultaneously and
democratically distributed into the active and sterile channels.

Atmospheric neutrino oscillations are governed by an effective $2\nu$
hamiltonian with vacuum mass-mixing parameters $(m^2,\psi)$, and with an
effective fermion density in matter given by $N_f=\frac{1}{2} s^2_\xi N_n$
[Eq.~(\ref{calHA})].  The case of negative $m^2$ is equivalent to swap the
first two octants of $\psi$, and thus it does not need a separate treatment.

Solar neutrino oscillations are governed by an effective $2\nu$ hamiltonian
with vacuum mass-mixing parameters $(\delta m^2,\omega)$, and with an effective
fermion density in matter given by $N_f=N_e-\frac{1}{2} c^2_\xi N_n$
[Eq.~(\ref{calHS})].  The case of negative $\delta m^2$ is equivalent to swap
the first two octants of $\omega$, and thus it does not need a separate
treatment.

The sign of $M^2$ (namely, the occurrence of a  solar $\nu$ doublet heavier or
lighter  than the atmospheric $\nu$ doublet) is irrelevant under our
approximations, as far as current oscillation experiments are concerned. Such
sign can instead be relevant in other contexts, such as in supernovae (see,
e.g., \cite{SNOV}) in big-bang nucleosynthesis (BBN) (see, e.g.,
\cite{YBBN,GBBN} and refs.\ therein)  and, for Majorana neutrinos, in
neutrinoless double beta decay ($0\nu2\beta$) (see, e.g., \cite{Bbet,Sbet,Kbet}
and refs.\ therein). In particular, BBN data can  probe new (sterile) neutrino
species in addition to the three active neutrinos, and thus constrain $4\nu$
scenarios. However,  at present, such constraints do not appear to be stable
enough  \cite{Lisi} to really exclude specific $4\nu$ models with great
confidence, especially if one elaborates upon a possible tension between the
most recent BBN and cosmic microwave background data \cite{Espo}. Moreover,
subtle effects of sterile neutrino oscillations in the early universe
\cite{Foot}, which are  still subject to intensive studies (see \cite{DiBa} and
refs.\ therein) might provide ways to evade or mitigate standard BBN bounds.
Constraints from $0\nu2\beta$ decay can also be evaded by assuming Dirac
neutrinos.  For such reasons, we postpone the discussion of (potentially
important but presently uncertain) additional BBN and $0\nu 2\beta$ bounds to a
future work, while in this paper we prefer to focus on a homogeneous set of
neutrino oscillation data (SK atmospheric).

Finally, we point out that the reduction from the general $4\nu$ oscillation
Hamiltonian (\ref{H'}) to  effective $2\nu$ forms  for both atmospheric and
solar neutrinos [Eqs.~(\ref{calHA}) and (\ref{calHS}), respectively] justifies
{\em a posteriori\/} our neglect of CP violation phases (unobservable in $2\nu$
scenarios), and thus the initial position (\ref{Ureal}).  An alternative proof
of the vanishing of (observable) CP violation effects in our framework can also
be obtained by applying our approximations (\ref{gerarchia}), (\ref{Umu12}) and
(\ref{Ue34})  to  the $4\nu$ Jarlskog invariants explictly worked out in
\cite{BaCP}.

\section{Graphical representations}

In our $4\nu$ framework, it turns out that the mixing parameter spaces for
atmospheric and solar neutrinos can be usefully represented in  triangular
plots, embedding unitarity relations  of the kind $U^2_1+U^2_2+U^2_3=1$ through
the heights projected  within a triangle with equal sides (and unit total
height).%
\footnote{Triangle plots have been already introduced and discussed in detail
in the context of $3\nu$ mixing, see \protect\cite{F3nu,Tria}.}
Such unitarity relations hold for the four columns of the mixing matrix $U$ 
characterizing our framework [Eq.~(\ref{Usimeq})],  and can be separated into
two constraints on the atmospheric $(\nu_3,\nu_4)$ doublet,
\begin{eqnarray}
U^2_{\mu3} + U^2_{\tau 3} + U^2_{s 3} &=& 1\ ,
\label{U3}\\
U^2_{\mu4} + U^2_{\tau 4} + U^2_{s 4} &=& 1\ ,
\label{U4}
\end{eqnarray}
and two constraints on the solar $(\nu_1,\nu_2)$ doublet,
\begin{eqnarray}
U^2_{e1} + U^2_{\tau 1} + U^2_{s 1} &=& 1\ ,
\label{U1}\\
U^2_{e2} + U^2_{\tau 2} + U^2_{s 2} &=& 1\ .
\label{U2}
\end{eqnarray}
As we will see below, it is sufficient to implement  just  one unitarity
relation in one triangle plot  for each doublet, the other relation being a
consequence. We choose to implement (\ref{U4}) for atmospheric neutrinos and
(\ref{U2}) for solar neutrinos.

Figure~2 represents the ``atmospheric neutrino triangle,''  embedding the
unitarity constraint (\ref{U4}), which is related to the flavor composition of
$\nu_4$,
\begin{eqnarray}
\nu_4 &=& U_{\mu 4}\,\nu_\mu + U_{\tau 4}\,\nu_\tau + U_{s 4}\,\nu_s 
\label{U41}\\
&=& s_\psi\,\nu_\mu+c_\psi c_\xi\,\nu_\tau+c_\psi s_\xi\, \nu_s
\label{U42}\\
&=& s_\psi\, \nu_\mu + c_\psi\, \nu_+\ .
\label{U43}
\end{eqnarray}
In the triangle plot, the upper, lower left and lower right corner are
identified with $\nu_\mu$, $\nu_\tau$, and $\nu_s$, respectively. The heights
projected from a generic point $\nu_4$ inside the triangle  onto the lower,
right, and left side  represent the elements $U^2_{\mu4}$, $U^2_{\tau 4}$, and
$U^2_{s 4}$, respectively.  When $\nu_4$ coincides with one of the corners 
(mass eigenstate = flavor eigenstate), no oscillation occurs.  Generic inner
points describe mixed (active+sterile) atmospheric neutrino oscillations,
smoothly interpolating from pure $\nu_\mu\to\nu_\tau$ (left side) to pure
$\nu_\mu\to\nu_s$ (right side). In the upper triangle plot of Fig.~2,  the
$(s^2_\psi,s^2_\xi)$ parametrization is also charted.  The lower triangle plot
shows explicitly that $\nu_4$ is a linear combination of $\nu_\mu$ and $\nu_+$,
with $\nu_+$ confined to the lower side (being a linear combination of
$\nu_\tau$ and $\nu_s$).  Notice that, once the point
$\nu_4=(s^2_\xi,s^2_\psi)$ is fixed, the position of $\nu_3$ is also determined
in the same plot  at the coordinates $(s^2_\xi,c^2_\psi)$, and there is no need
for a separate triangle embedding the constraint (\ref{U3}). The points $\nu_3$
and $\nu_4$ are then symmetrically placed onto the line joining the states
$\nu_\mu$ and $\nu_+$ (of which they are orthogonal combinations).

Figure~3 represents the ``solar neutrino triangle,'' embedding the unitarity
constraint (\ref{U2}),  which is related to the flavor composition of $\nu_2$,
\begin{eqnarray}
\nu_2 &=& U_{e 2}\,\nu_e + U_{\tau 2}\,\nu_\tau + U_{s 2}\,\nu_s 
\label{U21}\\
&=& s_\omega\,\nu_e-c_\omega s_\xi\,\nu_\tau+c_\omega s_\xi\, \nu_s
\label{U22}\\
&=& s_\omega\, \nu_e + c_\omega\, \nu_-\ .
\label{U23}
\end{eqnarray}
In the triangle plot, the upper, lower left and lower right corner are
identified with $\nu_e$, $\nu_s$, and $\nu_\tau$, respectively. The heights
projected from a generic point $\nu_2$ inside the triangle  onto the lower,
right, and left side  represent the elements $U^2_{e2}$, $U^2_{s 4}$, and
$U^2_{\tau 4}$, respectively.  When $\nu_2$ coincides with one of the corners 
(mass eigenstate = flavor eigenstate), no oscillation occur.  Generic inner
points describe mixed (active+sterile) solar neutrino oscillations, smoothly
interpolating from pure $\nu_\mu\to s$ (left side) to pure $\nu_\mu\to\nu_\tau$
(right side). In the upper triangle plot of Fig~3,  the $(s^2_\psi,s^2_\omega)$
parametrization is also charted.  The lower triangle plot shows explicitly that
$\nu_2$ is a linear combination of $\nu_e$ and $\nu_-$, with $\nu_-$ confined
to the lower side (being a linear combination of $\nu_s$ and $\nu_\tau$).
Notice that, once the point $\nu_2=(s^2_\xi,s^2_\omega)$ is fixed, the position
of $\nu_1$ is also determined in the same plot  with coordinates
$(s^2_\xi,c^2_\omega)$, and there is no need for a separate triangle embedding
the constraint (\ref{U1}). The points $\nu_1$ and $\nu_2$ are then
symmetrically placed onto the line joining the states $\nu_e$ and $\nu_-$ (of
which they are orthogonal combinations).

Finally, Fig.~4 shows the link between the solar and atmospheric parameter
spaces, which follows from the fact that the variable $\sin^2\xi$ must
represent the same abscissa in both triangles of Figs.~2 and 3.  By putting the
two triangles  on top of each other, it follows that the $\nu_+$ and $\nu_-$
points must be placed on the same vertical line, and that  couple of  points
inside the atmospheric and solar triangles  must be placed on iso-$s^2_\xi$
lines (slanted dashed lines in Fig.~4). It will be shown that such ``common
$\xi$'' constraint will significantly reduce the solutions that would be
separately allowed by solar data only or by atmospheric data only.

\section{Super-Kamiokande atmospheric $\nu$ data and expectations}

In the previous sections we have set the theoretical framework, as well as the
graphical representations, for our analysis of $4\nu$ oscillation solutions to
the atmospheric $\nu$ anomaly. In this section we present the data used in the
fit, and make a first discussion  (to be refined later) of the relative
importance of different data samples in constraining the mixing of atmospheric
$\nu_\mu$ with $\nu_s$.

Figure~5 shows the Super-Kamiokande data used in our $\chi^2$ analysis, as
graphically reduced from \cite{So2K}, corresponding to a detector exposure of
70.5 kTy. The reported errors are statistical only ($\pm 1\sigma$). Systematic
errors are treated as in \cite{F3nu}. The data are shown as binned
distributions of $e$-like and $\mu$-like event rates $R$,  normalized to 
no-oscillation expectations $R_0$ in each bin, in terms of the lepton zenith
angle $\vartheta$.  The distributions refer to sub-GeV (SG) leptons, multi-GeV
(MG) leptons, upward stopping (US) muons, and  upward through-going (UT) muons,
according to the SK terminology \cite{So2K,F3nu}. The total number of data
points used in the fit is 55.  The SK data on neutral-current enriched samples
\cite{SK2K}, which are also useful in testing the presence of a $\nu_s$
component \cite{Neut}, are not included in the present analysis.

Although our main results are based on accurate calculations \cite{F3nu} of the
full zenith distributions in Fig.~5, it is useful to gain some prior
understanding of the physics by means of  (partly) averaged quantities such as
the popular $\mu/e$ double flavor ratio for SG and MG events, the up-down
asymmetry  of MG events, and the vertical-to-horizontal ratio of UT events. We
follow the SK conventions for such quantities \cite{So2K,SK2K}, namely, data
with $|\cos\vartheta|<0.2$ are removed in the up-down asymmetry, and the 
separation between ``vertical'' and  ``horizontal'' through-going muons is
conventionally taken at $\cos\vartheta=-0.4$.

Figure~6 shows the double ratio $R_{\mu/e}$  of MG events {\em vs\/} SG events.
The corresponding SK data are shown as a cross (with statistical and total
$1\sigma$ error bars). The theoretical expectations at maximal mixing are shown
as a solid  line for pure $\nu_\mu\to\nu_\tau$ oscillations ($s^2_\xi=0$ in our
notation) and as a dashed line  for pure $\nu_\mu\to\nu_s$ oscillations 
($s^2_\xi=1$). Such lines connect the points calculated at $\pm m^2=n\times
10^{m}$ eV$^2$, with $m=-5,\,-4,\dots,\,-1$, and $n=1,\,2,\dots,\,9$ (the sign
of $m^2$ is irrelevant at maximal mixing, $s^2_\psi=1/2$). Cases with 
$0<s^2_\xi<1$ (not shown) would give lines intermediate between the solid and
the dashed ones. The ``trajectories'' of the theoretical points start at
$R_{\mu/e}\simeq 1$ for small $m^2$ (no-oscillation limit), and end at
$R_{\mu/e}\simeq 1/2$ for large $m^2$ (averaged oscillations).

From Fig.~6 we learn that the double ratio is not a good variable to
discriminate between $\nu_\mu\to\nu_\tau$ and $\nu_\mu\to\nu_s$ oscillations,
since the separation of the two corresponding curves is much smaller than the
experimental error. In fact, the matter effects which should distinguish the
$\nu_s$ channel are suppressed both by the  relatively low energy of SG and MG
events and by integration over the zenith angle.  On the other hand,  the
combined  SG and MG information on $R_{\mu/e}$ in Fig.~6 is sensitive to $
m^2$, and favors the range  $\sim [10^{-3},10^{-2}]$ eV$^2$,  with $\sim
4\times 10^{-3}$ eV$^2$ being the value closest to the data for both
oscillation channels. We conclude that the SG and MG $\mu/e$ double ratios give
valuable information on $m^2$, rather than on the flavor of the $\nu_\mu$
oscillation partner.

A better sensitivity to the sterile component of the $\nu_\mu$ oscillation
partner can be gained by improving the  zenith angle information and by probing
higher energies, where matter effects are larger \cite{Matt}. Figure~7 shows
the ratio of vertical to horizontal ($V/H$) upward through-going muons  {\em
vs\/} the up-down asymmetry of multi-GeV muons $(U-D/U+D)$.  As in Fig.~6,
solid (dashed) lines refer to $\nu_\mu\to\nu_\tau$ ($\nu_\mu\to\nu_s$)
oscillations at maximal mixing, $s^2_\psi=1/2$. In addition, Fig.~7 shows, as a
dotted line, the case of mixed active-sterile oscillations with equal
amplitude  [$\nu_+=(\nu_s+\nu_\tau)/\sqrt{2}$, namely, $s^2_\xi=1/2$]. The
trajectories of the theoretical points start and end at small values of the
up-down asymmetry, as expected in the limits $m^2\to 0$ (no oscillation) and
$m^2\to\infty$ (averaged oscillations), after passing through negative values
(corresponding to a  suppression of the upgoing muon rate). The $V/H$ ratio
should also  take back its no-oscillation value as $m^2\to\infty$,  although
the highest values reported in Fig.~7 ($m^2 = 10^{-1}$ eV$^2$) is not 
``asymptotic'' enough to show this behavior. The $V/H$ variations with $m^2$
are rather small in the pure $\nu_\mu\to\nu_s$ case, when compared with the
corresponding variations in the pure $\nu_\mu\to\nu_\tau$ case, as a
consequence of the strong damping of the sterile transition amplitude in matter
at high energies \cite{Matt}. Intermediate variations of $V/H$ occur for the
mixed active+sterile case (dotted line); therefore, the $V/H$ asymmetry is a
sensitive probe of $s^2_\xi$ through the effects of the matter term
$\sqrt{2}\,G_F\,s^2_\xi\,N_n/2$.

The recent SK data \cite{So2K}, shown in Fig.~7 as a cross with error bars
(statistical and total at $1\sigma$) clearly favor the pure
$\nu_\mu\to\nu_\tau$ case (solid line),  as compared with the pure
$\nu_\mu\to\nu_s$ case (dashed line). By themselves, such data cannot exclude
an intermediate situation with large mixing of $\nu_\mu$ with $\nu_\tau$ {\em
and\/} $\nu_s$ (dotted line). In such case, however, the favored values of
$m^2$ are rather large [$\sim O( 10^{-2}$] eV$^2$), in contrast with the
information coming from the $\mu/e$ ratio in Fig.~6. Therefore, the interplay
of different pieces of data seems to indicate clearly an upper bound on the
$\nu_s$ component of $\nu_+$ (namely,  on $s^2_\xi$).

\section{Results of the atmospheric $\nu$ analysis}

In this section we present the results of our fit to the atmospheric neutrino
data of Fig.~5 (55 data points),  as obtained by calculating the SK zenith
distributions for unconstrained values of the three relevant parameters
$(m^2,s^2_\xi,s^2_\psi)$. We discuss first the bounds on the mass parameter
$m^2$, and then those on the mixing parameters $(s^2_\xi,s^2_\psi)$.

Figure~8 shows the value of the global  $\chi^2$ as a function of $m^2$ (in
linear scale),  for four $\nu_\mu\to\nu_+$ oscillation cases, corresponding to:
$(i)$ unconstrained $\nu_+$ (thick solid line); $(ii)$ $\nu_+=\nu_\tau$ (thin
solid line); $(iii)$ $\nu_+=\nu_s$ (dashed line); and $(iv)$
$\nu_+=(\nu_s+\nu_\tau)/\sqrt{2}$ (dotted line).  In terms of the mixing angle
$\xi$, such cases correspond, respectively, to: $(i)$ unconstrained $s^2_\xi$;
$(ii)$ $s^2_\xi=0$;  $(iii)$ $s^2_\xi=1$; and $(iv)$ $s^2_\xi=1/2$. The
parameter $s^2_\psi$ is left free in all four cases.  The global minimum for
the unconstrained case $(\chi^2_{\min}=47.4)$ is reached at  $(m^2/{\rm
eV}^2,s^2_\xi,s^2_\psi)=(3.2\times 10^{-3},0.18,0.51)$. The slight preference
for a nonzero value of $s^2_\xi$ at  the point of minimum $\chi^2$ is
intriguing, but not statistically significant, since the fit with $s^2_\xi=0$
(pure $\nu_\mu\to\nu_\tau$ oscillations, thin solid line)  is only slightly
worse than the unconstrained fit (thick solid line).

Concerning the statistical interpretation of the results shown in Fig.~8,  some
remarks are in order. In general, one can use the $\chi^2$ function for two
complementary purposes \cite{PDGR}:  (1) to estimate the best-fit values of
$(m^2,s^2_\xi,s^2_\psi)$ and their uncertainties ({\em parameter estimation});
or (2) to test  the goodness-of-fit with some prior assumptions about
$(m^2,s^2_\xi,s^2_\psi)$ ({\em hypothesis test}). In the first approach, the 
values of $(m^2,s^2_\xi,s^2_\psi)$ minimizing the $\chi^2$ function are taken
as best estimates of the (unknown) true values of such parameters. The
corresponding uncertainties are then obtained by allowing a  shift in the
$\chi^2$ function ($\Delta\chi^2=\chi^2-\chi^2_{\rm min}$), with corresponding
confidence levels determined by the number of free  parameters in the model
($N_{\rm DF}=3$ in our case). From this point of view, the case of pure
$\nu_\mu\to\nu_s$ oscillations, corresponding to $\Delta\chi^2/N_{\rm
DF}=14.7/3$ in Fig.~8, is reached only at 99.8\% C.L., and is thus strongly
disfavored. In the second approach,  one checks how well a specific hypothesis
about the parameters  fits the 55 SK data points, independently of the
existence of a global $\chi^2$ minimum at some point in the parameter space.  
For instance, if order to test the {\em a priori\/} hypotheses of pure
$\nu_\mu\to\nu_\tau$ or pure $\nu_\mu\to\nu_s$ oscillations, one should compare
the absolute $\chi^2$ value with $N_{\rm DF}=55-2$, the two free variables
being $m^2$ and $s^2_\psi$ ($s^2_\xi$ being fixed at 0 or 1, respectively). 
From Fig.~8, we then get that the probability of having a worse $\chi^2$ is
63\% for $s^2_\xi=0$ (pure $\nu_\mu\to\nu_\tau$ oscillations, $\chi^2/N_{\rm
DF}=49/53$)  and 18\% for $s^2_\xi=1$ ($\chi^2/N_{\rm DF}=49/53$ (pure
$\nu_\mu\to\nu_s$ oscillations, $\chi^2/N_{\rm DF}=62/53$). Therefore, although
the pure sterile case is clearly disfavored {\em in comparison\/} with the pure
active case, it is not ruled out {\em a priori\/} in our oscillation analysis. 
Having clarified such statistical issues, we proceed to estimate  the
mass-mixing parameters $(m^2,s^2_\xi,s^2_\psi)$ by using the first approach, 
based on $\Delta\chi^2$.

In Fig.~8, all four cases show a preference  for $m^2 \simeq 3\times 10^{-3}$
eV$^2$, which thus represents a very stable indication  coming from the SK
data, independently of the amplitude of the $\nu_s$ component. For the most
general case (unconstrained $\nu_+$) we obtain
\begin{equation}
m^2 \simeq 3.2^{+2.3}_{-1.6}\times 10^{-3}\ {\rm eV}^2\  (90\% {\rm\ C.L.})\ ,
\end{equation}
by taking $\Delta\chi^2=6.25$ for $N_{\rm DF}=3$.

Figure~8 also gives gives indications about the  parameter $s^2_\xi$. The case
$s^2_\xi=0$ (thin solid line) gives results comparable to the best fit for
unconstrained $s^2_\xi$ (thick solid line). The case of a fifty-fifty admixture
of $\nu_s$ and $\nu_\tau$ in $\nu_+$ ($s^2_\xi=1/2$, dotted line)   provides a
fit that, although worse than for pure $\nu_\mu\to\nu_\tau$,  is still 
acceptable (its minimum $\chi^2$ being only four units above the absolute
minimum). However, the quality of the fit worsen rapidly  as $\nu_+$ gets
closer to $\nu_s$, the worst case being reached for pure $\nu_\mu\to\nu_s$
oscillations (dashed line). Therefore, we expect to get an upper bound on 
$s^2_\xi$ from the analysis of the mixing parameters, which can be  clearly
represented in the triangle plot.

Figures~9 and 10 show the results of our analysis in the  atmospheric triangle
plot, for $10$ representative (decreasing) values of $m^2$. The first three
columns of triangles refer to the separate fits  to sub-GeV electrons and muons
(10+10 bins), multi-GeV electrons and muons (10+10 bins), and upward stopping
and through-going muons (5+10 bins), while the fourth column refers to the
total SK data sample (55 bins). For each column, we find the minimum $\chi^2$
from the fit to the corresponding data sample, and then present sections of the
allowed volume at fixed values of $m^2$ for  $\Delta\chi^2=6.25$ (90\% C.L.,
solid lines) and $\Delta\chi^2=11.36$ (99\% C.L., dotted lines). It can be seen
that the low-energy SG data are basically insensitive to $s^2_\xi$, since they
are consistent both with $\nu_\mu\to\nu_\tau$ oscillations (left side) and with
$\nu_\mu\to\nu_s$ oscillations (right side), as well as with any intermediate
combination of the two  oscillation channels. High-energy upgoing $\mu$ data are
instead much more sensitive to the $\nu_s$ component through matter effects,
which increase both with energy and with $s^2_\xi$, and tend to suppress the
$\mu$ deficit \cite{Matt}. Large matter effects appear to be  disfavored by the
upgoing muon data,   the rightmost part of the triangle being excluded at 90\%
C.L. The multi-GeV data cover an intermediate energy range and are not as
constraining as the upgoing muons; however, they also  show a tendency to
disfavor a large $\nu_s$ component. This tendency is strengthened in the 
global combination of data (fourth column), leading to the upper bound
\begin{equation}
s^2_\xi \lesssim 0.67\  (90\% {\rm\ C.L.})\ ,
\label{xibound}
\end{equation}
which clearly disfavors pure or quasi-pure $\nu_\mu\to\nu_s$ oscillations. In
particular, from our Figs.~9 and 10 we derive that pure $\nu_\mu\to\nu_s$
oscillations ($s^2_\xi=1$) are excluded at $>99\%$ C.L.,  consistently with the
SK analysis \cite{So2K,SK2K}.

However, the bound (\ref{xibound}) still allows intermediate cases of combined
active+sterile oscillations with a sizable $\nu_s$ component. For instance,
one cannot exclude that the sterile neutrino channel may have the same
amplitude as the active one $(s^2_\xi=0.5)$, or may even be dominant  (e.g.,
$s^2_\xi=0.6$, corresponding to  $\nu_+ \simeq 0.77\,\nu_s+0.65\,\nu_\tau$).
Furthermore, for $m^2$ in its upper range (e.g., $m^2=5\times 10^{-3}$ eV$^2$
in Fig.~9), a subdominant $\nu_s$ component actually helps in the global fit to
SK data, leading to a 90\% C.L. allowed region which does not touch the left
side of the triangle (corresponding to the pure $\nu_\mu\to\nu_\tau$ channel).

Figures~9 and 10 also provide bounds on the  other mixing parameter $s^2_\psi$,
which governs the relative amount of $\nu_\mu$ and $\nu_+$ on the atmospheric
neutrino states $\nu_3$ and $\nu_4$ [see Eqs.~(\ref{numu},\ref{nu++}) and
Fig.~2], and thus the overall amplitude of $\nu_\mu\to\nu_+$ oscillations. From
Figs.~9 and 10 we derive
\begin{equation}
s^2_\psi \simeq 0.51\pm 0.17\  (90\% {\rm\ C.L.})\ ,
\end{equation}
which favors $\nu_\mu\to\nu_+$ oscillations with nearly maximal amplitude, as
one expects from the observation of nearly maximal average suppression ($\sim
50\%$) of upgoing MG muons. There is a slight asymmetry of the allowed regions
with respect to $s^2_\psi=1/2$, which reflects the relative  change of sign of
vacuum and matter terms in Eq.~(\ref{calHA}) when passing from the first to the
second octant.

Figures~8, 9, and 10 demonstrate that, on the basis of present SK data on the
zenith distributions of leptons induced by atmospheric neutrinos, pure
$\nu_\mu\to\nu_s$ oscillations are strongly  disfavored, but one cannot exclude
mixed active-sterile $\nu_\mu\to\nu_+$ oscillations with
$\nu_+=c_\xi\,\nu_\tau+s_\xi\,\nu_s$,  provided that the partial amplitude of
the sterile channel is $\lesssim 67\%$ [Eq.~(\ref{xibound})].  Since such bound
is dominated by high-energy muons, it is important to understand how such data
can improve the sensitivity to $s^2_\xi$ in the future.

Figure~11 shows the zenith distributions of upward through-going muons at
maximal $(\nu_\mu,\nu_+)$ mixing $(s^2_\psi=1/2)$ for $\nu_+=\nu_\tau$ (left
panel), $\nu_+=(\nu_\tau+\nu_s)/\sqrt{2}$ (middle panel), and $\nu_+=\nu_s$
(right panel). The solid (dashed) histograms refer to $\pm m^2=3\times 10^{-3}$
eV$^2$  ($5\times 10^{-3}$ eV$^2$), the sign of $m^2$  being irrelevant at
maximal mixing. From left to right, $s^2_\xi$ increases and matter effects also
grow with it, producing an increasing modulation of the zenith distributions.
The modulation, which is very strong for the oscillation probability at
fixed energies (not shown), is largely smeared out in Fig.~11 by integration
over the energy spectrum and, to a lesser extent, by the presence of both $\nu$
and $\overline \nu$ (with opposite matter terms) in the atmospheric neutrino
flux. Therefore, the modulation pattern of the  UT$\mu$ distribution might be
more clearly observed in future atmospheric $\nu$ experiments \cite{Ge2K}
aiming at improved energy reconstruction and $\mu^+/\mu^-$ separation.
Concerning the present SK data (Fig.~11), the large scatter of central values
and the still large error bars prevent a clear discrimination of the
theoretical distributions in the left and middle panels, which anyway appear to
provide a better fit than in right panel. A reduction by a factor $\lesssim 2$
in the uncertainties of the SK UT$\mu$ sample appears necessary for a decisive
discrimination among the three cases shown in Fig.~11. In conclusion, we can
still learn a lot on the $\nu_s$ component of $\nu_+$ as we get more
high-energy muon data in SK, as well as in MACRO, and possibly from future
atmospheric $\nu$ experiments.

\section{Linking atmospheric and solar neutrinos}

Solar neutrino oscillations with both active and sterile states have been
recently investigated in \cite{Pena,Pe2K}, whose results can be interpreted, in
our framework, through the identification $c^2_{23}c^2_{24}\equiv 1-s^2_\xi$,
$\theta_{12}=\omega$, and $\Delta m^2_{21}=\delta m^2$.  In \cite{Pena,Pe2K} it
has been shown that there is a continuous set of solutions to the solar
neutrino problem for unconstrained $s^2_\xi$, ranging from pure ``active
oscillation solutions'' ($s^2_\xi=1$) to pure ``sterile oscillation solutions''
($s^2_\xi=0$). In particular, for $s^2_\xi=1$  all the usual solutions to the
solar neutrino problem are present, either at small $\omega$ mixing 
(matter-enhanced solution) or at large $\omega$ mixing (matter-enhanced
solution at low or high $\delta m^2$, or vacuum solutions). For decreasing
values of $s^2_\xi$, all the solutions tend to shrink; however,  the large
mixing (LM) ones (vacuum or matter-enhanced) eventually disappear for
$s^2_\xi\lesssim0.3$--$0.4$, while the small mixing (SM) solution still
survives (at $\sim 99\%$ C.L.) even for $s^2_\xi=0$.  Such behavior is
qualitatively reported in Fig.~12, through the triangular representation of the
parameter space. The plot is only qualitative, since the detailed behavior of
any solution depends both on the confidence level adopted and on the specific
value chosen for $\delta m^2$, whose complete scan is beyond the scope of this
paper.

In Fig.~12, the allowed region at large mixing ($s^2_\omega$ close to $1/2$),
which has the maximum likelihood at $s^2_\xi=1$ (pure active oscillations,
right side of the triangle), gradually shrinks when $s^2_\xi$ decreases and
eventually disappears. The allowed region at small mixing ($s^2_\omega$ close
to 0) covers instead the whole $s^2_\xi$ range.

As discussed in Secs.~II and III,  the solar and atmospheric solutions are
coupled through the common angle $\xi$. Such coupling can be visualized by
putting the solar and atmospheric triangles on top of each other, as shown in
Fig.~4. The results are  qualitatively different for the SM and LM solutions to
the solar neutrino problem, as shown in Figs.~13 and 14.

Figure~13 shows, in grey, the regions globally allowed by the combination of
solar and atmospheric $\nu$ data, assuming the SM solution to the solar
$\nu$ problem. The fact that atmospheric $\nu$ data put an upper bound on 
$s^2_\xi$ (as quantitatively discussed through Figs.~9 and 10) implies that
also the fraction of the SM solar $\nu$ solution at large $s^2_\xi$ is
excluded. However, there is still a continuous range of global solutions for
$s^2_\xi\neq 0$, having a sizable $\nu_s$ component for both solar and
atmospheric $\nu$ oscillations. Therefore atmospheric $\nu$ data ``cut'' a
fraction  of the SM solution to the solar $\nu$ problem, but the combination
still allows solutions involving active+sterile neutrino oscillations for both
atmospheric and solar neutrinos.

The opposite happens for the LM solution to the solar neutrino problem. In this
case, solar neutrino data cut the ``small $s^2_\xi$''  part of the atmospheric
allowed region, as shown in Fig.~14, and the globally allowed solutions (grey
inner regions) represent genuine active+sterile oscillations, which never
reduce to the pure active or pure sterile subcases (the sides of the
triangles). In particular, the points denoted by stars in Fig.~14 represent a
peculiar solution (which might be dubbed ``fourfold maximal mixing''),
corresponding to maximal mixing of $\nu_{\pm}$ with $\nu_{s,\tau}$
($s^2_\xi=1/2$), maximal mixing of  $\nu_{3,4}$ with $\nu_{\mu,+}$
($s^2_\psi=1/2$), and maximal mixing of $\nu_{1,2}$ with $\nu_{e,-}$
($s^2_\omega=1/2$). In this case,  both solar and atmospheric neutrino
oscillations are equally distributed among the active and sterile channels.

In conclusion, it turns out that, at least qualitatively, $4\nu$ scenarios with
mixed active+sterile transitions for both solar and atmospheric  neutrinos are
allowed by the combination of world oscillation data, with different mixing
patterns dictated by the solution assumed for the solar neutrino problem 
(either small or large mixing). Consistency between atmospheric and solar
neutrino data in $4\nu$ models has also been recently found  in an independent
and somewhat different analysis \cite{Ya00}.  Therefore, the hypothesis of a
fourth, sterile neutrino is still compatible with present data and needs
further theoretical and experimental investigations. Such results clearly
demonstrate that the two-flavor approximations often used to analyze solar or
atmospheric neutrinos data (in terms of either pure active or pure sterile
oscillations) are, in general, not appropriate to explore the full parameter
space allowed by $4\nu$ models. Our $4\nu$ formalism and triangular
representations can also be usefully applied to explore the  $\nu_\mu\to\nu_s$
discovery potential in long baseline accelerator experiments, as shown for the
MINOS experiment in \cite{Pety}.

\section{Remarks on 3+1 mass spectra}

The standard arguments disfavoring 3+1 spectra (triplet+singlet) with respect
to 2+2 spectra (two doublets) in the interpretation of the  current $4\nu$
oscillation phenomenology \cite{Four} appear to be somewhat weakened
\cite{Fate,Sm2K} in light of the most recent LSND data \cite{LS2K}, and should
be perhaps revisited quantitatively. Here we make just a few qualitative
comments on the mixing pattern that is likely to emerge in 3+1 models
characterized by a triplet $(\nu_1,\nu_2,\nu_3)$ (mainly responsible for solar
and atmospheric oscillations) and by a ``loner'' state $\nu_4$ (mainly
responsible for LSND oscillations).

The absence (or weak occurrence, as in LSND) of oscillations in all channels
probed by short-baseline reactor and accelerator experiments is consistent with
$\nu_4$ being close to a flavor eigenstate. The dominant flavor of $\nu_4$
cannot be $\nu_e$  ($\nu_\mu$), otherwise  the disappearance pattern of solar
$\nu_e$ (atmospheric $\nu_\mu$) would remain unexplained. Moreover, if it were
$\nu_4\sim \nu_\tau$, then the triplet $(\nu_1,\nu_2,\nu_3)$ would
approximately be  a linear combination of $(\nu_\mu,\nu_e,\nu_s)$, and thus 
the atmospheric $\nu_\mu$ oscillation partner should be either $\nu_e$
(excluded) or $\nu_s$ (disfavored at $\gtrsim 99\%$ C.L.) or a linear
combination of $\nu_e$ and $\nu_s$ (presumably disfavored). Therefore, the
favored option for 3+1 spectra seems to imply $\nu_4\sim \nu_s$, with (at
least) a small subdominant component of $\nu_\mu$ and $\nu_e$ in $\nu_4$, 
necessary to drive LSND oscillations (see also \cite{Fate,PNOW}).%
\footnote{Alternative 3+1 models with small $(\nu_s,\nu_4)$ mixing are
currently being investigated, see \protect\cite{GNOW}.}

Assuming $\nu_4\sim \nu_s$ in 3+1 models,  the triplet  states
$(\nu_1,\nu_2,\nu_3)$ would then approximately be  linear combinations of
$(\nu_e,\nu_\mu,\nu_\tau)$, and the triplet phenomenology of solar and
atmospheric neutrinos would practically become equivalent to that of a standard
framework with three active neutrinos. Such $3\nu$ framework has been worked
out in detail in previous papers \cite{F3nu,Tria,3neu},  where the reader can
find the corresponding constraints on the mass-mixing parameters. In
conclusion, it seems that the emerging mixing pattern  for the 3+1 spectrum
implies, up to small ``LSND perturbations'', an effective three-neutrino
phenomenology for solar and atmospheric oscillations. In this case, it becomes
important to study how well one can experimentally discriminate such $4\nu$
option from pure $3\nu$ oscillations.

\section{Summary and conclusions}

In the context of 2+2 neutrino models, we have thoroughly analyzed the SK
atmospheric data on the zenith distributions of lepton events, assuming the
coexistence of both $\nu_\mu\to\nu_\tau$ and $\nu_\mu\to\nu_s$ oscillations,
with a smooth interpolation between such two subcases. We have shown that,
although the data disfavor oscillations in the pure $\nu_\mu\to\nu_s$ channel,
one cannot exclude their presence {\em in addition\/} to $\nu_\mu\to\nu_\tau$  
oscillations. High energy muon data appear to be crucial to assess the relative
amplitude of the active and sterile oscillation channels for atmospheric
neutrinos.  We have also shown, in a qualitative way, that mixed active+sterile
oscillations of atmospheric neutrinos are compatible with analogous
oscillations of solar neutrinos, with different mixing pattern emerging from
different choices for the solar $\nu$ solutions.  Such results indicate that
the solar and atmospheric neutrino phenomenology cannot be simply embedded in
the usual two-family scenarios involving oscillations into either pure active
or pure sterile states, and call for an extended $4\nu$ framework involving
combined solutions. Further directions to improve and constrain our present
results might involve, on the one hand, a quantitative, global analysis of {\em
world\/} neutrino oscillation data and, on the other hand, the inclusion of
non-oscillatory  data which have not been considered in this work, such as
those coming from  astrophysics and cosmology, and from direct neutrino mass
measurements.

\acknowledgments

E.L.\ thanks the organizers of the 19th International Conference on Neutrino 
Physics and Astrophysics (Sudbury, Canada), where preliminary results of this
work were presented, for kind hospitality. We thank 
C.\ Giunti, 
R.\ Foot,
T.\ Kajita, 
D.\ Montanino, 
M.\ Messier,
R.\ Mohapatra,
D.\ Petyt,
F.\ Ronga,  
F.\ Vissani,
and O.\ Yasuda 
for useful discussions or comments. This work was co-financed by the Italian
Ministero dell'Universit\`a e della Ricerca Scientifica e Tecnologica  (MURST)
within the ``Astroparticle Physics'' project.


%
\newcommand{\InsertFigure}[2]{\newpage\begin{center}\mbox{%
\epsfig{bbllx=1.4truecm,bblly=1.3truecm,bburx=19.5truecm,bbury=26.5truecm,%
height=21.4truecm,figure=#1}}\end{center}\vspace*{-1.8truecm}%
\parbox[t]{\hsize}{\small\baselineskip=0.5truecm\hspace*{0.5truecm} #2}}
\InsertFigure{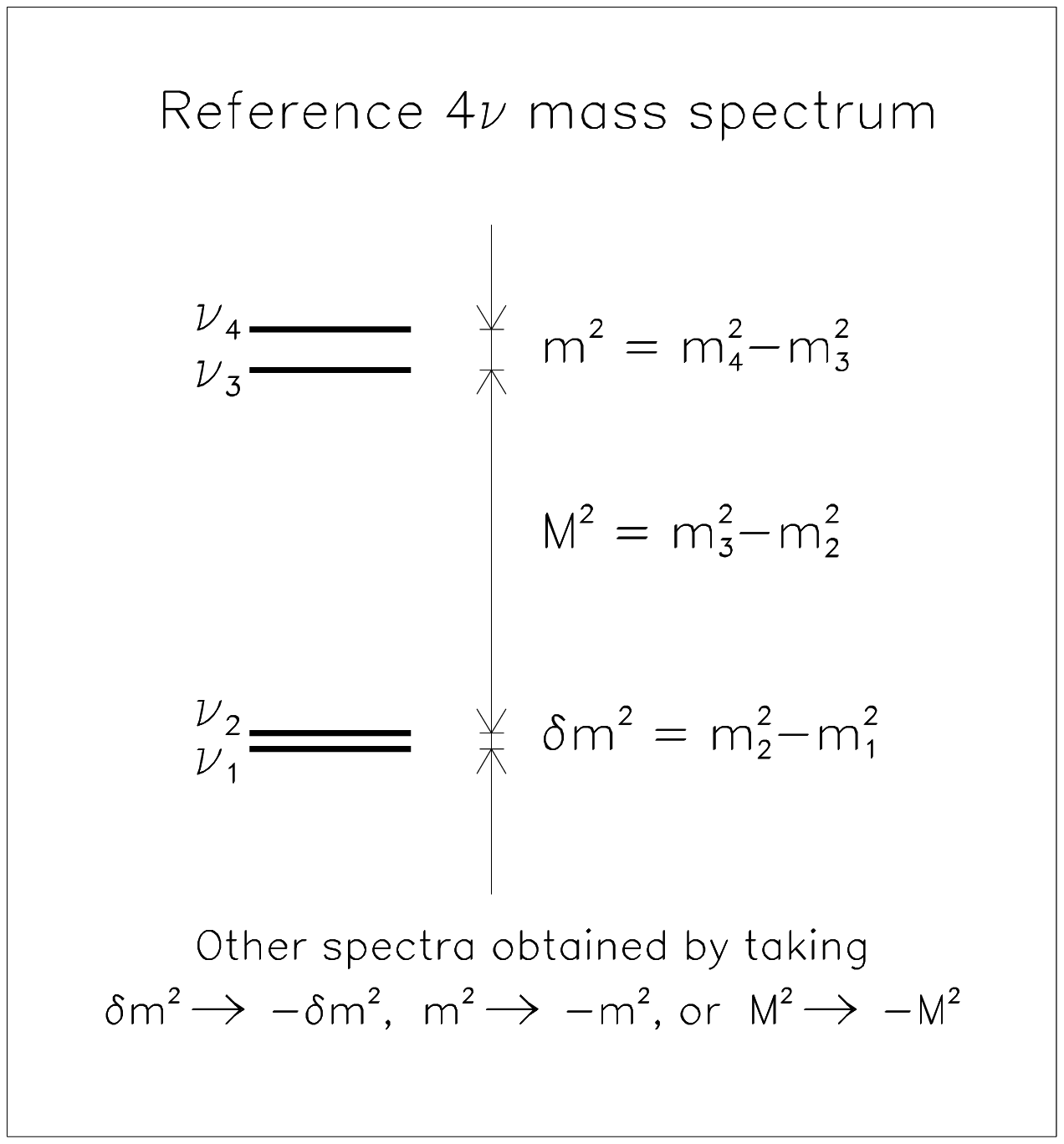}%
{Fig.~1. The reference $4\nu$ mass spectrum adopted in this work.}
\InsertFigure{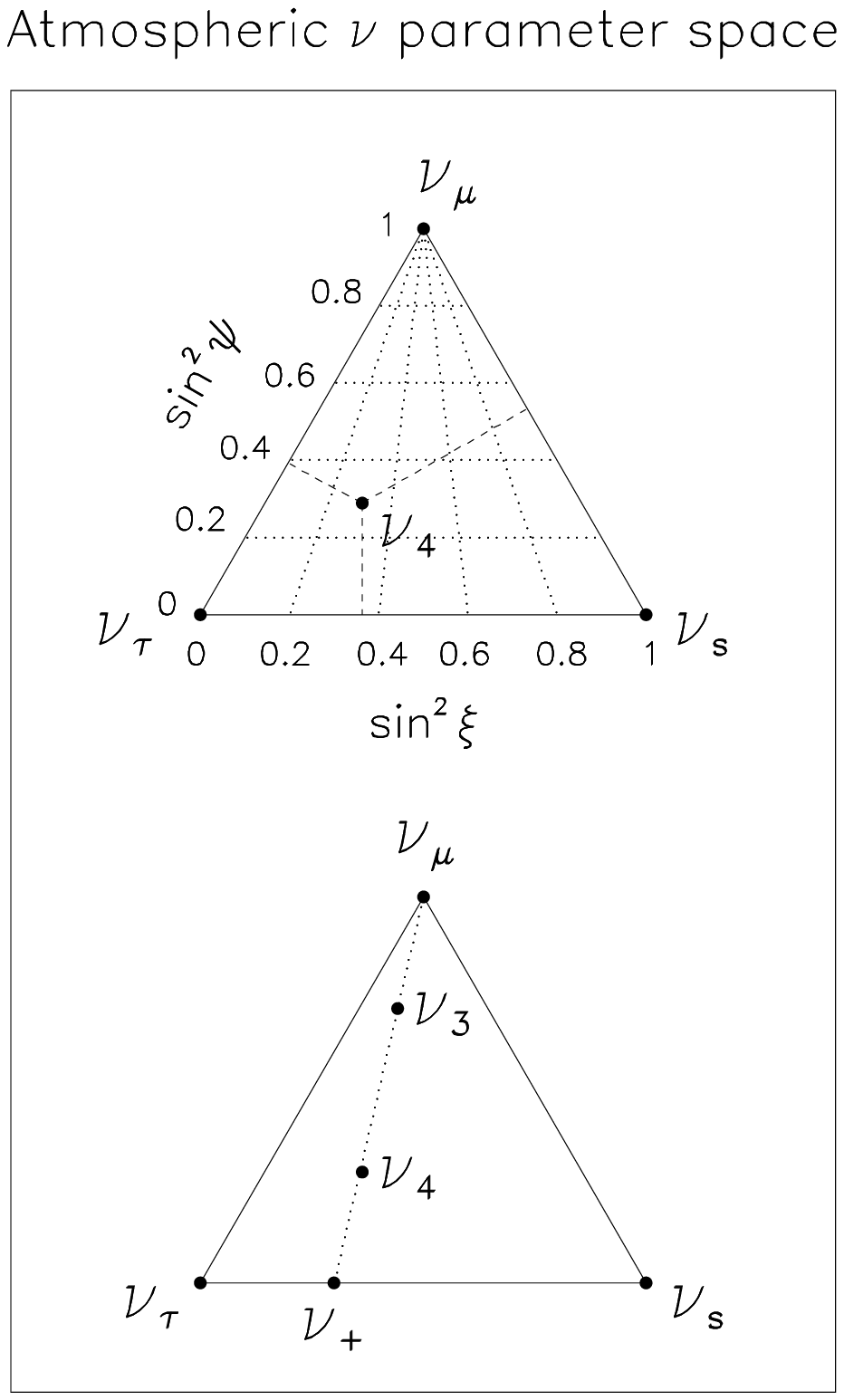}%
{Fig.~2. The atmospheric $\nu$ parameter space, in the triangular 
representation. See the text for details.}
\InsertFigure{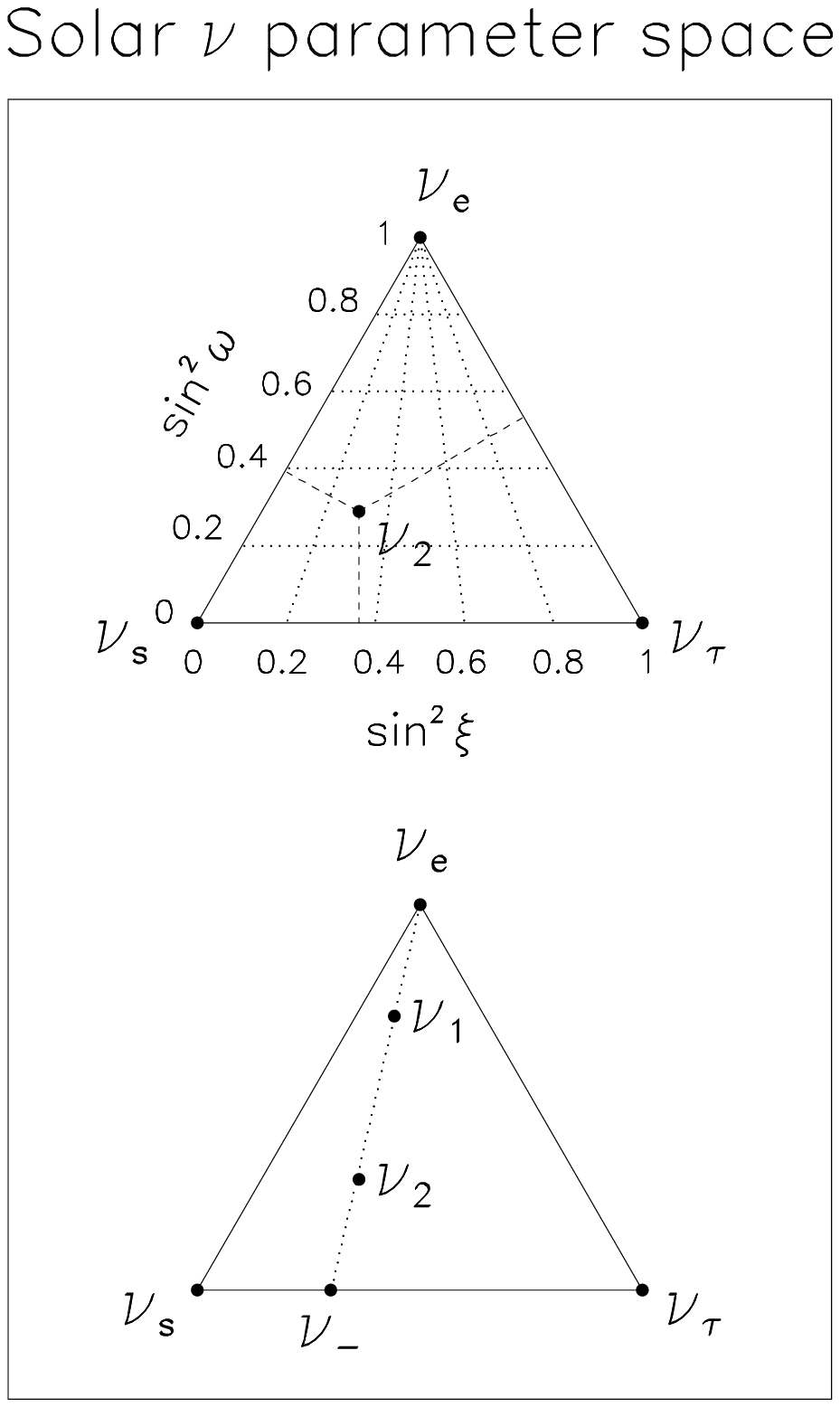}%
{Fig.~3. The solar $\nu$ parameter space, in the triangular representation. See
the text for details.}
\InsertFigure{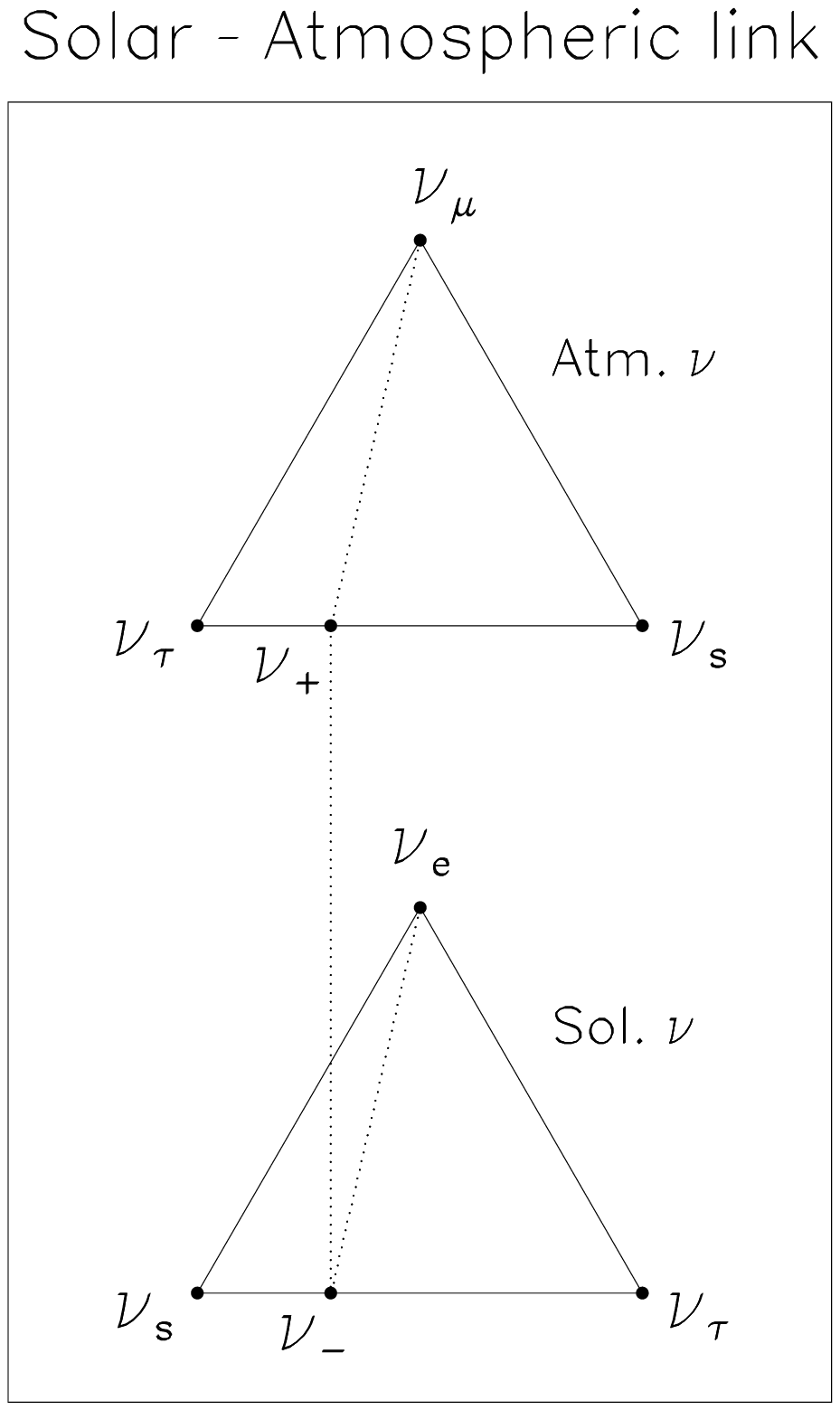}%
{Fig.~4. Graphical representation of the link existing between the solar and
atmospheric parameter spaces.}
\InsertFigure{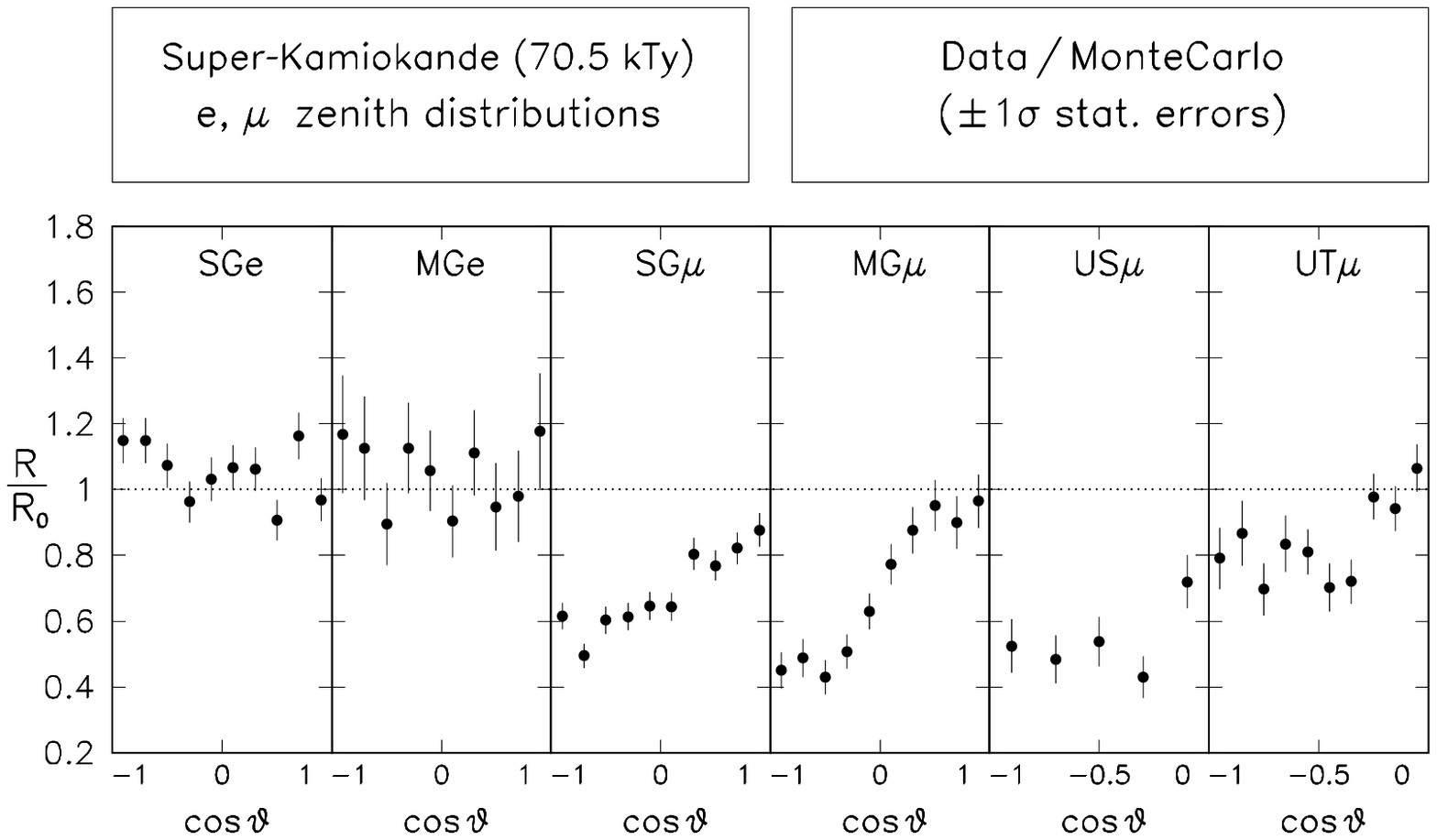}%
{Fig.~5. Super-Kamiokande data on the zenith distributions of lepton events
induced by atmospheric neutrinos.}
\InsertFigure{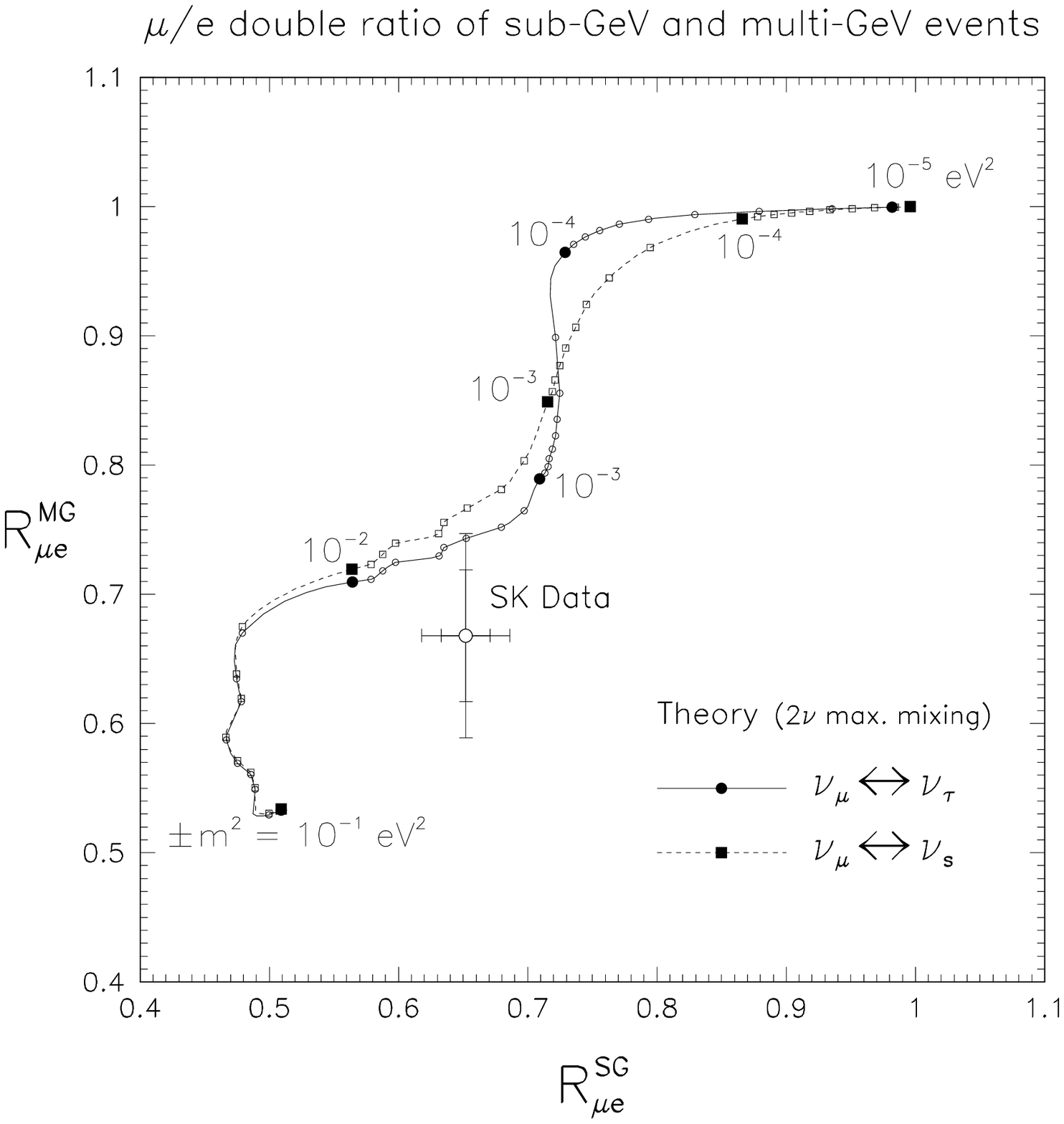}%
{Fig.~6. Double ratio of muon-to-electron events for multi-GeV events
($y$-axis) and sub-GeV events ($x$-axis).  The theoretical expectation at fixed
(maximal) mixing and for running $m^2$ is shown as a solid curve for pure
active oscillations, and as a dashed curve  for pure sterile oscillations. The
cross with error bars  (statistical and total) represents the SK data.}
\InsertFigure{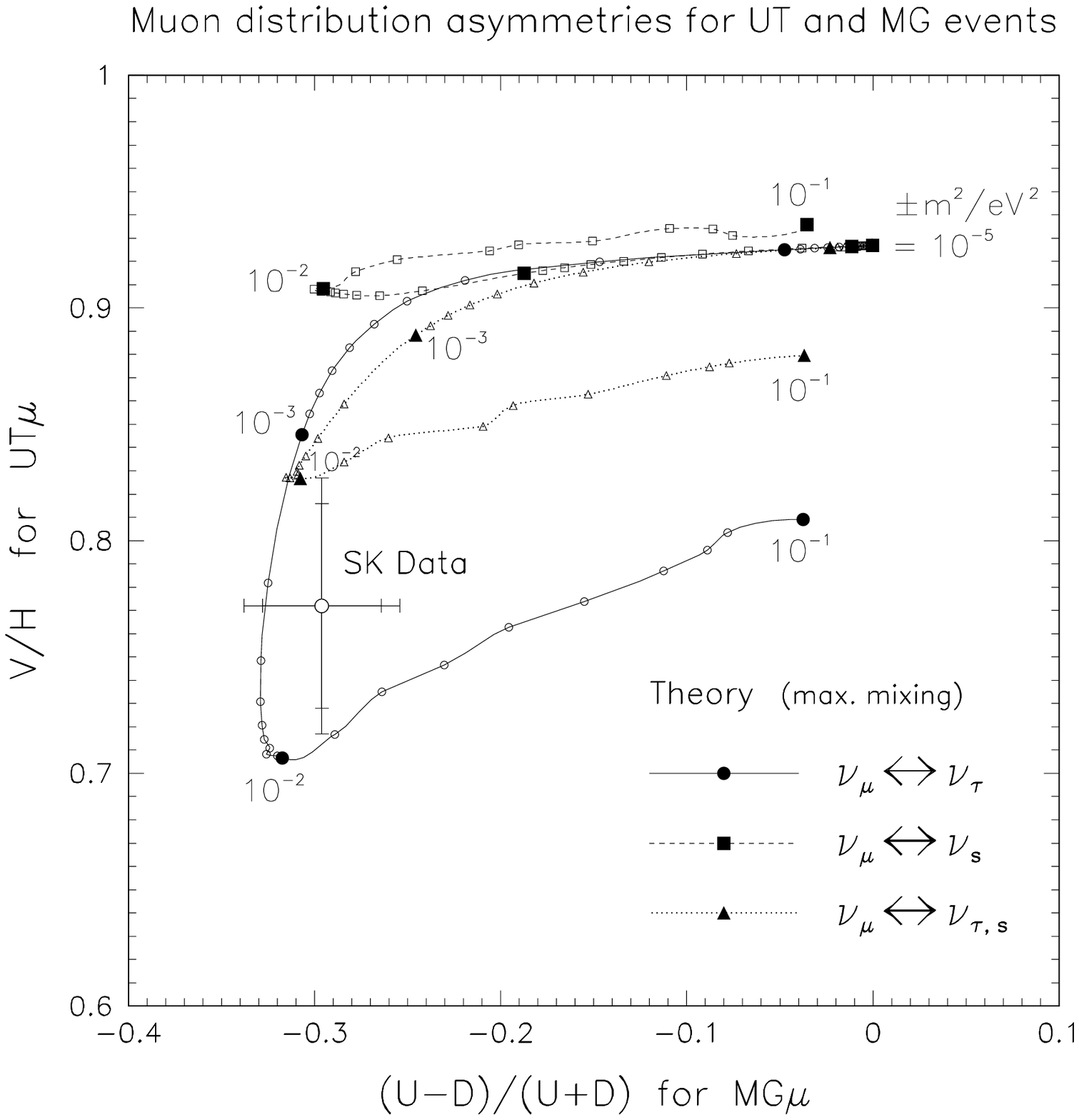}%
{Fig.~7. Vertical-to-horizontal ratio of upgoing muon rates {\em vs} the
up-down asymmetry of multi-GeV muon events.  The theoretical expectation at
fixed (maximal) mixing and for running $m^2$ is shown as a solid curve for pure
active oscillations, as a dashed curve  for pure sterile oscillations, and as a
dotted curve for mixed  active+sterile oscillations of equal amplitude. The
cross with error bars represents the SK data.}
\InsertFigure{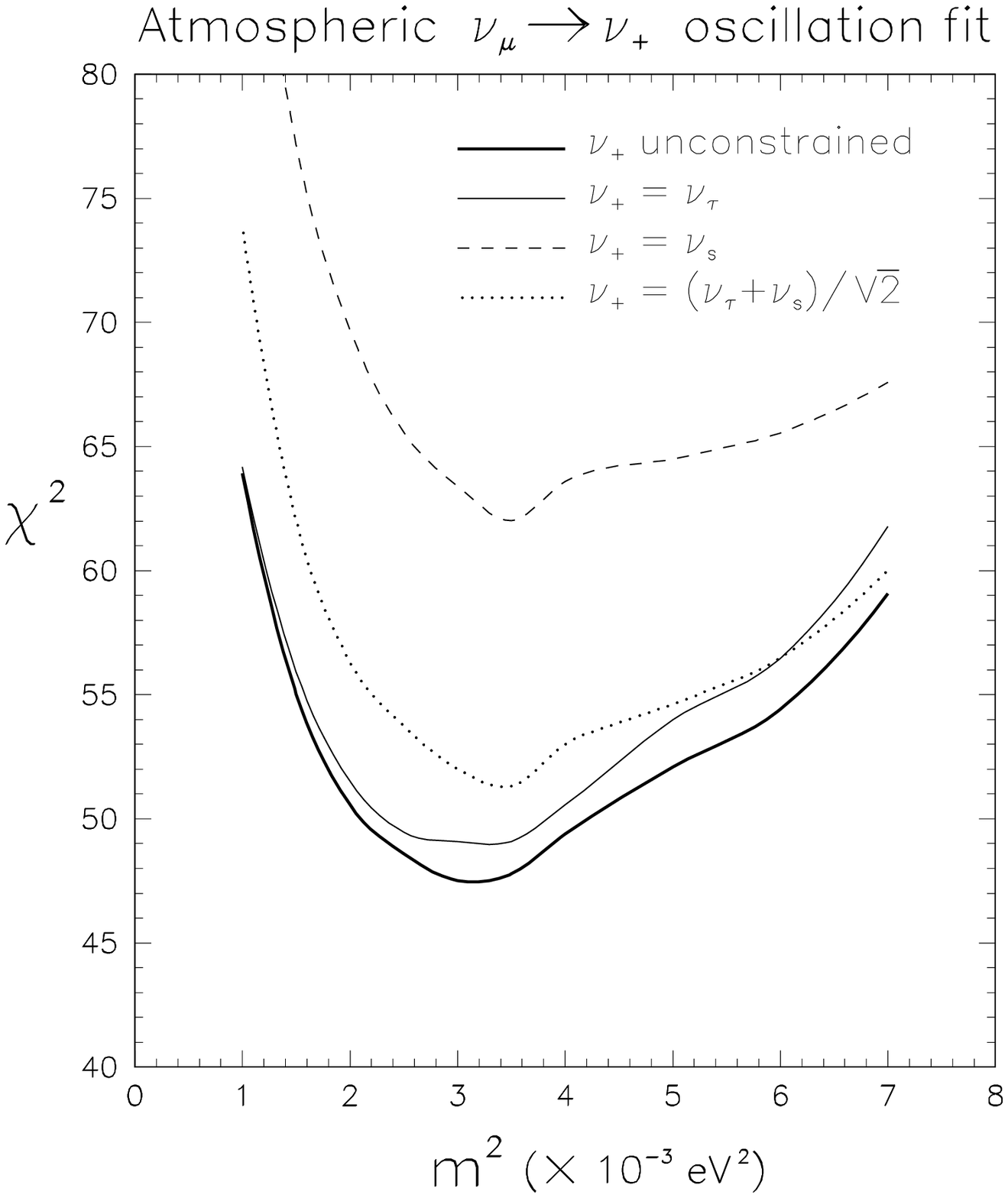}%
{Fig.~8. Result of the global $\chi^2$ fit to the 55 data points of Fig.~5, as
a function of the mass parameter $m^2$, assuming $\nu_\mu\to\nu_+$ oscillations
for atmospheric neutrinos, and different options for $\nu_+$. The mixing angle
$s^2_\psi$  is left unconstrained in all cases. The case $\nu_+=\nu_s$ provides
the worst fit, but one cannot exclude a sizable $\nu_s$ component of $\nu_+$. }
\InsertFigure{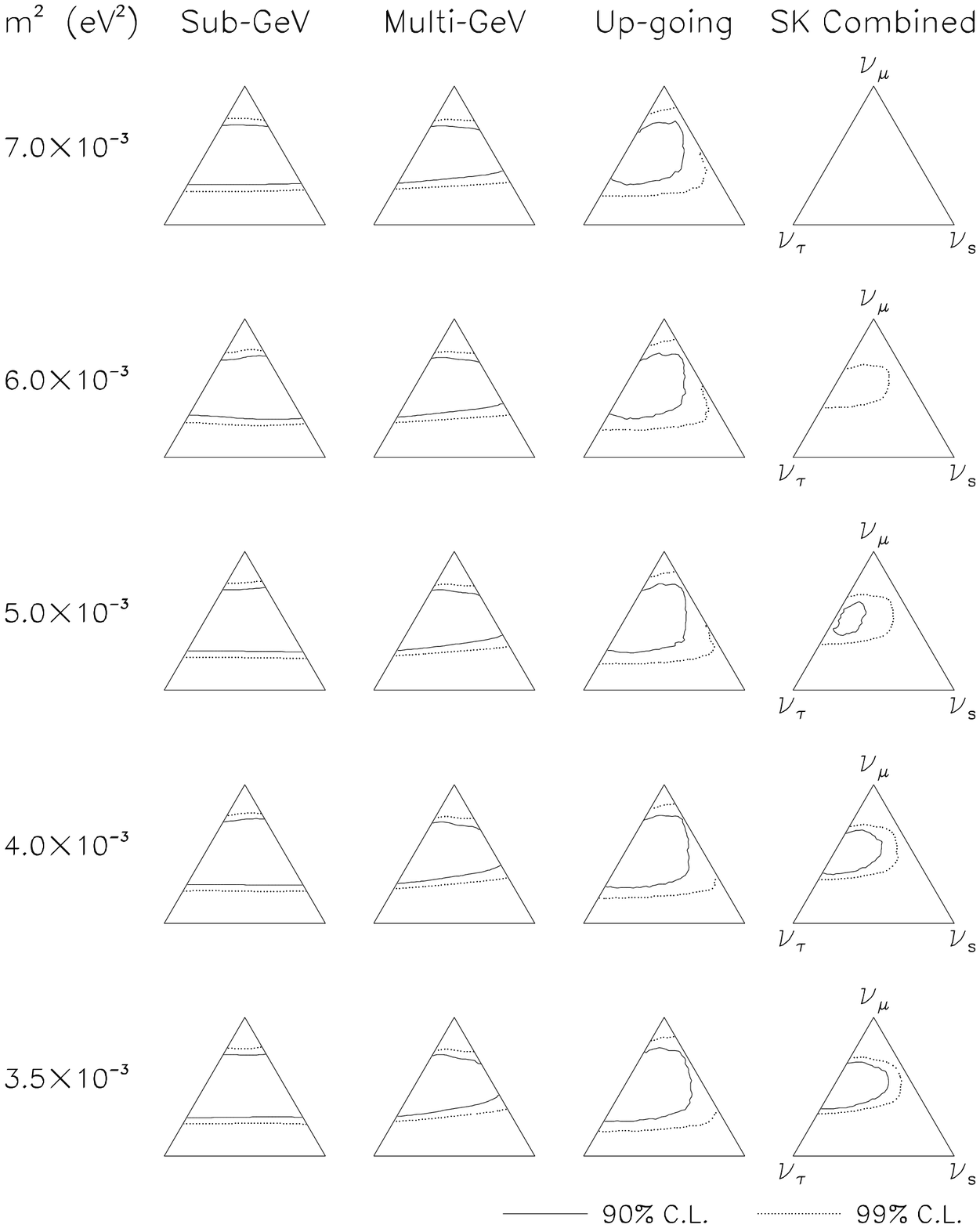}%
{Fig.~9. Results of the global $\chi^2$ fit to atmospheric neutrino data in the
triangular representation, for decreasing values of $m^2$. See the text for
details.}
\InsertFigure{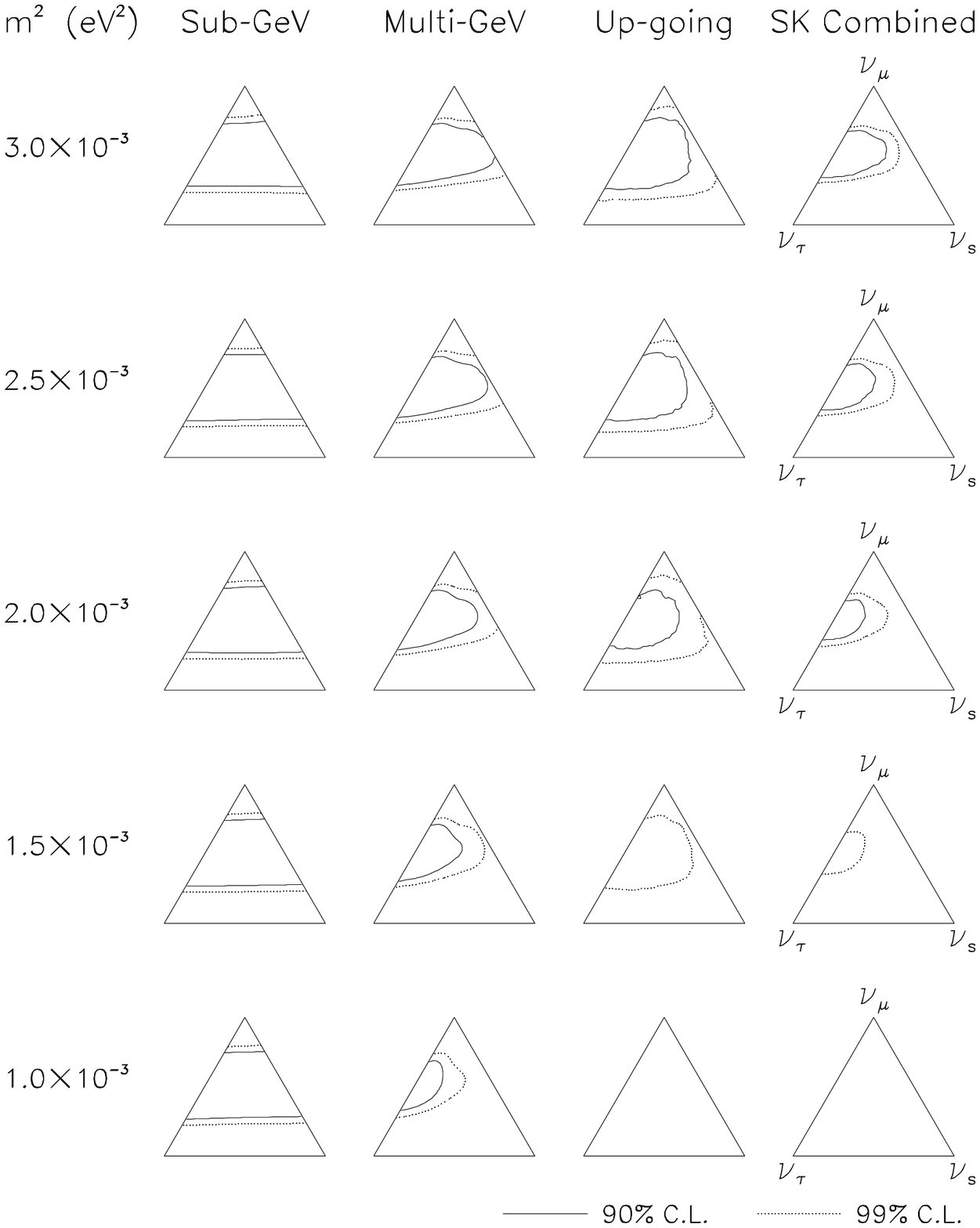}%
{Fig.~10. As in Fig.~9, but for lower values of $m^2$.}
\InsertFigure{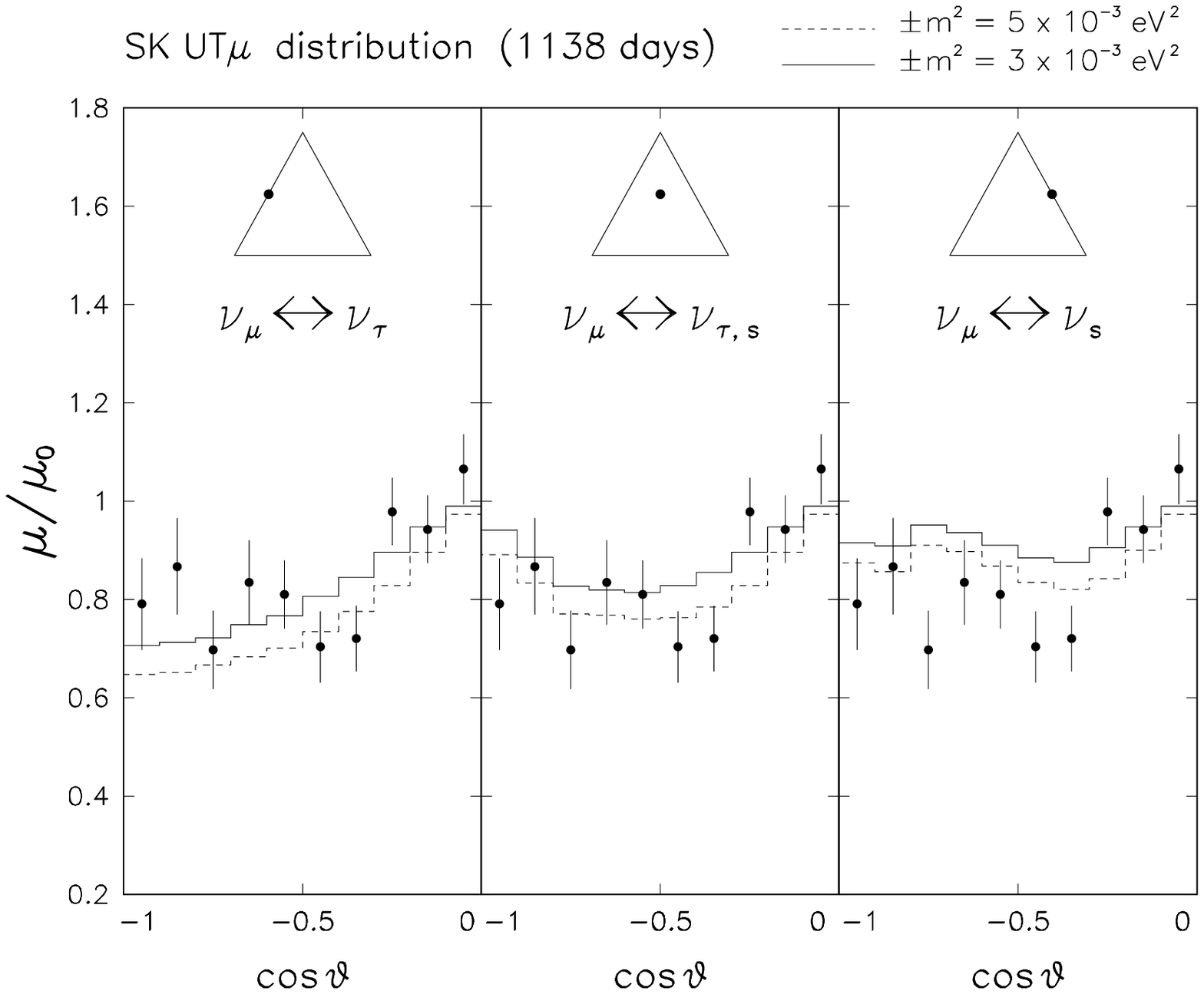}%
{Fig.~11. The zenith distribution of upward through-going muon events for three
different oscillation scenarios, involving pure active oscillations (left
panel), pure sterile oscillations (right panel) and mixed active+sterile
oscillations (middle panel) at maximal mixing.}
\InsertFigure{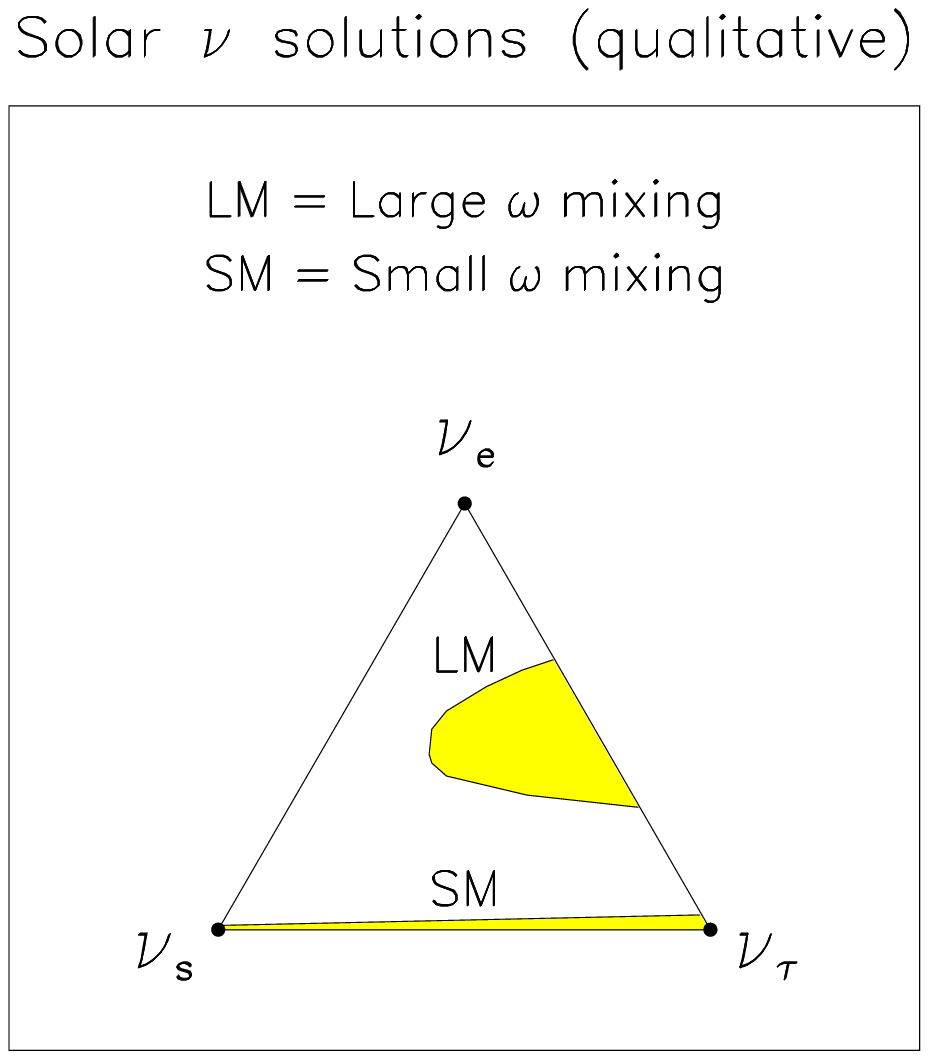}%
{Fig.~12. Qualitative pattern of $4\nu$ oscillation solutions to the solar
neutrino problem at small or large mixing, in the triangular representation of
the parameter space.}
\InsertFigure{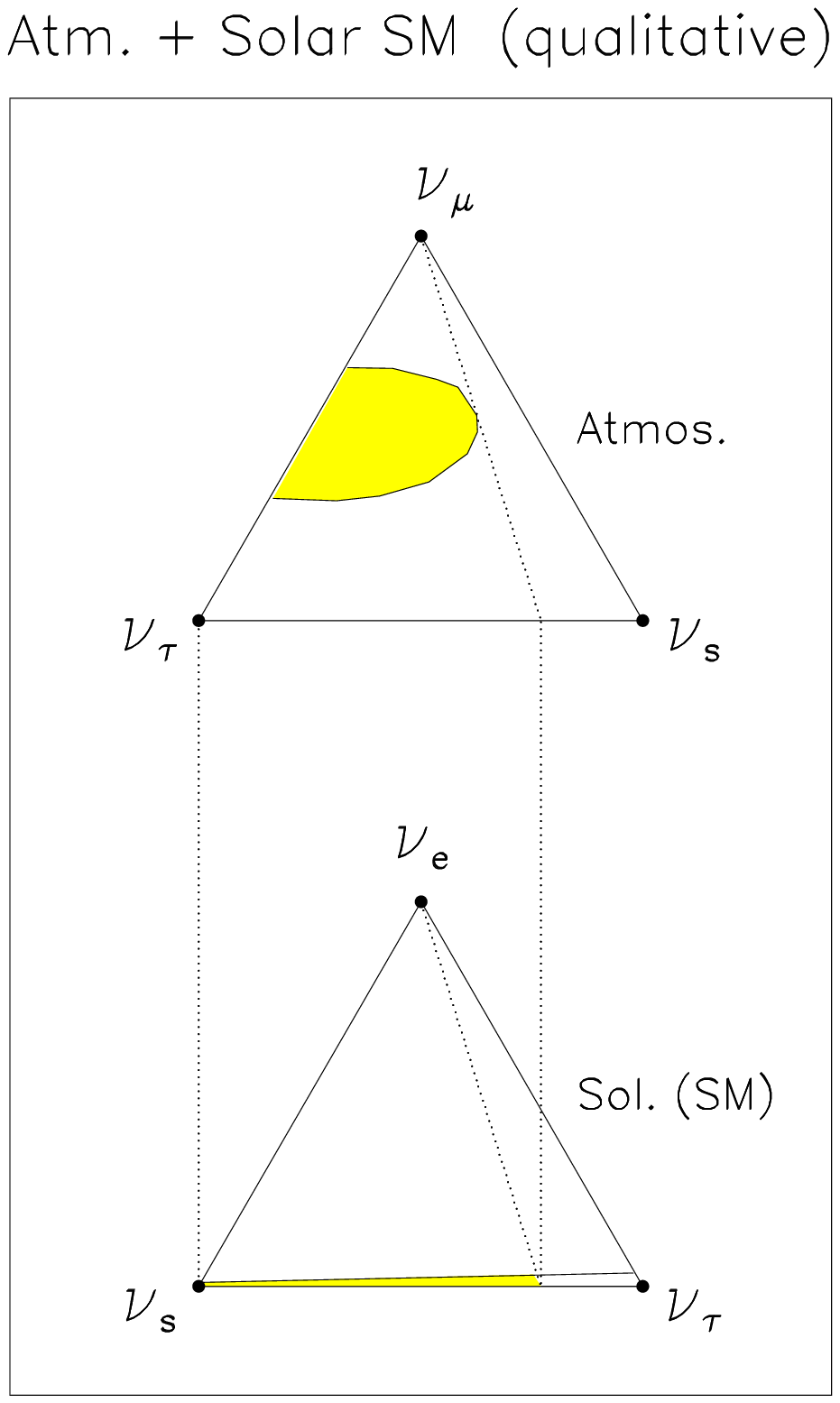}%
{Fig.~13. Qualitative combination of the atmospheric $\nu$ solution with a
typical small-angle solar $\nu$ solution.}
\InsertFigure{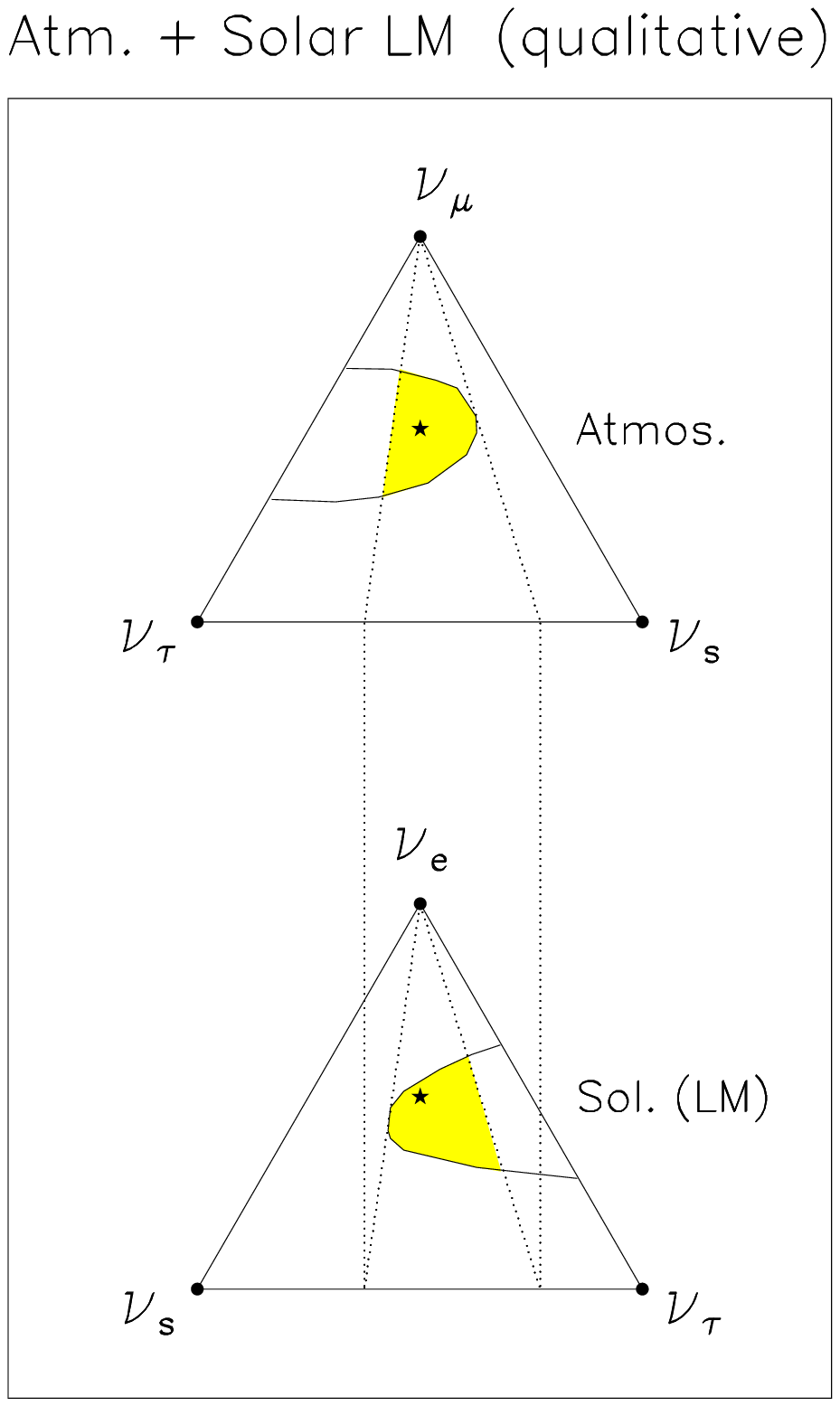}%
{Fig.~14. Qualitative combination of the atmospheric $\nu$ solution with a
typical large-angle solar $\nu$ solution.}

\eject

\begin{references}

\bibitem{Four}	S.M.\ Bilenky, C.\ Giunti, and W.\ Grimus,
		Proceedings of {\em Neutrino '96\/}, Helsinki, June 1996,
		edited by K.\ Enqvist {\em et al.} 
		(World Scientific, Singapore, 1997), p.174, hep-ph/9609343;
		V.\ Barger, S.\ Pakvasa, T.J.\ Weiler, and K.\ Whisnant,
		Phys.\ Rev.\ D {\bf 58}, 093016 (1998). For
		pre-LSND $2+2$ models, see also:
		D.O.\ Caldwell and R.N.\ Mohapatra, 
		Phys.\ Rev.\ D {\bf 48}, 3259 (1993); 
		J.T.\ Peltoniemi and J.W.F.\ Valle,
		Nucl.\ Phys.\ B {\bf 406}, 409 (1993).
		
\bibitem{Bi99}	S.M.\ Bilenky, C.\ Giunti, and W.\ Grimus,
		Prog.\ Part.\ Nucl.\ Phys.\ {\bf 43}, 1 (2000).
		
\bibitem{Do00}	D.\ Dooling, C.\ Giunti, K.\ Kang, and C.W.\ Kim,
		Phys.\ Rev.\ D {\bf 61}, 073011 (2000).	
		
\bibitem{Ba00}	V.\ Barger and K.\ Whisnant, hep-ph/0006235,
		in {\em Current Aspects of Neutrino Physics},
		ed.\ by D.\ Caldwell (Springer-Verlag, Hamburg, 2000).	
		
\bibitem{LSND}	LSND Collaboration, C.\ Athanassopoulos {\em et al.},
		Phys.\ Rev.\ Lett.\ {\bf 81}, 1774 (1998);
		Phys.\ Rev.\ C {\bf 58}, 2489 (1998).
					
\bibitem{LS2K}	G.\ Mills for the LSND Collaboration, 
		in {\em Neutrino 2000}, 
		19th International Conference on Neutrino 
		Physics and Astrophysics (Sudbury, Canada, 2000),
		to appear; transparencies available at the site 
		http://nu2000.sno.laurentian.ca
		
\bibitem{So2K}	H.\ Sobel for the Super-Kamiokande Collaboration,
		in {\em Neutrino 2000\/} \protect\cite{LS2K}.
		
\bibitem{SK2K}	Super-Kamiokande Collaboration, S.\ Fukuda {\em et al.},
		Phys.\ Rev.\ Lett.\ {\bf 85}, 3999 (2000).

\bibitem{Su2K}	Y.\ Suzuki for the Super-Kamiokande Collaboration,
		in {\em Neutrino 2000\/} \protect\cite{LS2K}.	

\bibitem{Ba2K}	B.\ Barish for the MACRO Collaboration,
		in {\em Neutrino 2000\/} \protect\cite{LS2K}.	
		
\bibitem{Go2K}	C.\ Gonzalez-Garcia, in {\em ICHEP 2000},
		30th International Conference on High Energy Physics
		(Osaka, Japan, 2000), to appear; transparencies available
		at the site  http://www.ichep2000.rl.ac.uk
		
\bibitem{Fate}	V.\ Barger, B.\ Kayser, J.\ Learned, T.\ Weiler,
		and K. Whisnant, Phys.\ Lett.\ B {\bf 489}, 345 (2000).
		
\bibitem{Pena}	C.\ Giunti, M.C.\ Gonzalez-Garcia, and 	C.\ Pe{\~n}a-Garay,
		Phys.\ Rev.\ D {\bf 62}, 013005 (2000). 

\bibitem{Pe2K}	For an update of the results discussed in 
		\protect\cite{Pena}, see 
		M.C.\ Gonzalez-Garcia and C.\ Pe{\~n}a-Garay,
		hep-ph/0009041.

\bibitem{Li2K}	G.L.\ Fogli, E.\ Lisi, and A.\ Marrone,   
		in {\em Neutrino 2000}	\protect\cite{LS2K}.	

\bibitem{Ya00}	O.\ Yasuda, hep-ph/0006319.
			
\bibitem{Pont}	B.\ Pontecorvo, 
		Zh.\ Eksp.\ Teor.\ Fiz.\ {\bf 53}, 1717 (1967)
		[Sov. Phys.\ JETP {\bf 26}, 984 (1968)];
		Z.\ Maki, M.\ Nakagawa, and S.\ Sakata,
		Prog.\ Theor.\ Phys.\ {\bf 28}, 675 (1962).
		
\bibitem{MSWs}	L.\ Wolfenstein, 
		Phys.\ Rev.\ D {\bf 17}, 2369 (1978);
		S.P.\ Mikheyev and A.\ Yu.\ Smirnov,
		Yad.\ Fiz.\ {\bf 42}, 1441 (1985) 
		[Sov.\ J.\ Nucl.\ Phys.\ {\bf 42}, 913 (1985)]; 
		Nuovo Cimento C {\bf 9}, 17 (1986);
		V.\ Barger, K.\ Whisnant, S.\ Pakvasa, and R.J.N.\ Phillips,
		Phys.\ Rev.\ D {\bf 22}, 2718 (1980).

\bibitem{Sm2K}	A.Yu.\ Smirnov,
		in {\em Neutrino 2000} \protect\cite{LS2K}.	
		
\bibitem{CDHS}	CDHSW Collaboration, F.\ Dydak {\em et al.}, 
		Phys.\ Lett.\ B {\bf 134}, 281 (1984).

\bibitem{SNOV}	D.O.\ Caldwell, G.M.\ Fuller, and Y.-Z.\ Qian,
		Phys.\ Rev.\ D {\bf 61}, 123005.

\bibitem{YBBN}	N.\ Okada and O.\ Yasuda,
		Int.\ J.\ Mod.\ Phys.\ A {\bf 12}, 3669 (1997).
		
\bibitem{GBBN}	S.M.\ Bilenky, C.\ Giunti, W.\ Grimus, and T.\ Schwetz,
		Astropart.\ Phys.\ {\bf 11}, 413 (1999).
		
\bibitem{Bbet}	S.M.\ Bilenky, C.\ Giunti, B.\ Kayser, and S.T.\ Petcov,
		Phys.\ Lett.\ B {\bf 465}, 193 (1999).
		
\bibitem{Sbet}	H.V.\ Klapdor-Kleingrothaus, H.\ P{\"a}s, and A.Yu.\ Smirnov,
		hep-ph/9910205, in {\em Beyond the Desert~'99}, Proceedings
		of the 2nd International Conference on Physics Beyond the 
		Standard Model: Accelerator, Nonaccelerator and Space
		Approaches (Tegernsee, Germany, 2000), edited by
		H.V.\ Klapdor-Kleingrothaus and I.V.\ Krivosheina
		(Bristol, IOP, 2000)
		
\bibitem{Kbet}	A.\ Kalliomaki and A.\ Maalampi,
		Phys.\ Lett.\ B {\bf 484}, 64 (2000).
		
\bibitem{Lisi}	E.\ Lisi, S.\ Sarkar, and F.L.\ Villante,
		Phys.\ Rev.\ D {\bf 59}, 123520 (1999).
		
\bibitem{Espo}	S.\ Esposito, G.\ Mangano, G.\ Miele, and O.\ Pisanti,
		J.\ of High Energy Phys.\ {\bf 9}, 38 (2000).

\bibitem{Foot}	R.\ Foot and R.R.\ Volkas, 
		Phys.\ Rev.\ D {\bf 55}, 5147 (1997).
		
\bibitem{DiBa}	P.\ Di Bari and R.\ Foot, hep-ph/0008258.
		
\bibitem{BaCP}	V.\ Barger, Y.B.\ Dau, K.\ Whisnant, and B.L.\ Young,
		Phys.\ Rev.\ D {\bf 59}, 113010 (1999).
		
\bibitem{F3nu}	G.L.\ Fogli, E.\ Lisi, A.\ Marrone, and G.\ Scioscia,
		Phys.\ Rev.\ D {\bf 59}, 033001 (1999).

\bibitem{Tria}	G.L.\ Fogli, E.\ Lisi, and D.\ Montanino, 
		Phys.\ Rev. D {\bf 54}, 2048 (1996).

\bibitem{Neut}	F.\ Vissani and A.Yu.\ Smirnov,
		Phys.\ Lett.\ B {\bf 432}, 376 (1998);
		L.J.\ Hall and H.\ Murayama,
		Phys.\ Lett.\ B {\bf 436}, 323 (1998).

\bibitem{Matt}	E.\ Akhmedov, P.\ Lipari, and M.\ Lusignoli,
		Phys.\ Lett.\ B {\bf 300}, 128 (1993);
		P.\ Lipari and M.\ Lusignoli,
		Phys.\ Rev.\ D {\bf 58}, 073005 (1998);
		Q.Y.\ Liu and A.Yu.\ Smirnov, 
		Nucl.\ Phys.\ B {\bf 524}, 505 (1998);
		Q.Y.\ Liu, S.P.\ Mikheyev, and A.Yu.\ Smirnov,
		Phys.\ Lett.\ B {\bf 440}, 319 (1998);
		R.\ Foot, R.\ Volkas, and O.\ Yasuda,
		Phys.\ Rev.\ D {\bf 58}, 013006 (1998).

\bibitem{PDGR}	Review of Particle Physics, D.E.\ Groom {\em et al.},
		Eur.\ Phys.\ J.\ C {\bf 15}, 1 (2000).

\bibitem{Ge2K}	A.\ Geiser, in {\em Neutrino 2000} \protect\cite{LS2K}.

\bibitem{Pety}	D.\ Petyt, ``$\nu_\mu\to\nu_s$ in MINOS,''
		MINOS internal note NuMI-L-691, available at 
		www.hep.anl.gov/ndk/hypertext/numi\_notes.html

\bibitem{PNOW}	O.L.G.\ Peres, talk at {\em NOW 2000}, 2nd Europhysics Neutrino
		Oscillation Workshop (Conca Specchiulla, Otranto, Lecce,
		Italy, 9-16 Sept.\ 2000), transparencies available
		at www.ba.infn.it/$^\sim$now2000;
		O.L.G.\ Peres and A.Yu.\ Smirnov,
		hep-ph/0011054.

\bibitem{GNOW}	C.\ Giunti, talk at {\em NOW 2000} \protect\cite{GNOW};
		C.\ Giunti and M.\ Laveder, hep-ph/0010009.

\bibitem{3neu}	G.L.\ Fogli, E.\ Lisi, D.\ Montanino, and
		G.\ Scioscia, Phys.\ Rev.\ D {\bf 55}, 4385 (1997);
		G.L.\ Fogli, E.\ Lisi, and A.\ Marrone,
		Phys.\ Rev.\ D {\bf 57}, 5893 (1998);
		G.L.\ Fogli, E.\ Lisi, D.\ Montanino, and A.\ Palazzo,
		Phys.\ Rev.\ D {\bf 62}, 013002 (2000);
		Phys.\ Rev.\ D {\bf 62}, 113004 (2000).

\end{references}
\end{document}